\documentclass[%
 reprint,
 amsmath,amssymb,
 aps,
prb,
]{revtex4-2}

\usepackage{graphicx}
\usepackage{dcolumn}
\usepackage{bm}
\usepackage{multirow}
\usepackage{siunitx}
\usepackage{hyperref}

\newcommand{\ket}[1]{\left| #1 \right\rangle}
\newcommand{\bra}[1]{\left\langle #1 \right|}

\newcommand{\Jex}{J_{\text{ex}}}

\newcommand{\lhf}{$\mathrm{LiHoF_4}$}
\newcommand{\lef}{$\mathrm{LiErF_4}$}

\newcommand{\lhfx}{$\mathrm{LiHo_{x}Y_{1-x}F_4}$}
\newcommand{\lhef}{$\mathrm{LiHo_{x}Er_{1-x}F_4}$}
\newcommand{\Ho}{$\mathrm{Ho^{3+}}$}

\newcommand{\Er}{$\mathrm{Er^{3+}}$}

\begin{document}

\preprint{APS/123-QED}

\title{Off-diagonal dipolar interactions in the mixed Ising--XY magnet  ${\mathbf{LiHo_{x}Er_{1-x}F_4}}$}

\author{Tomer Dollberg}
\author{Moshe Schechter}%
\affiliation{%
 Department of Physics, Ben-Gurion University of the Negev, Beer Sheva 84105, Israel
}%

\date{\today}

\begin{abstract}
We theoretically investigate the influence of off-diagonal dipolar interactions in the mixed-anisotropy magnet \lhef{}. Motivated by experimental observations showing an unexpectedly rapid suppression of the ferromagnetic transition temperature $T_c$ upon substitution of Ho by Er ions, we use Monte Carlo simulations incorporating off-diagonal dipolar terms to elucidate the underlying physical mechanism. Our results reveal that, unlike in the diluted magnet \lhfx{}, where dilution weakens the effect of these interactions, substitution by planar Er ions amplifies it. This leads to a pronounced reduction of $T_c$, closely matching experimental data, thereby providing a natural explanation for discrepancies with mean-field predictions. The findings underscore the essential role of off-diagonal dipolar interactions in determining the magnetic properties of mixed-anisotropy dipolar systems.

\end{abstract}

\maketitle

\section{Introduction}
Materials hosting distinct types of magnetic anisotropy offer diverse avenues for emergent and exotic magnetic behavior. In particular, mixed or random anisotropy magnets can exhibit rich phase diagrams and unconventional ordering behaviors due to the interplay between competing spin orientations and frustration effects \cite{albenRandomAnisotropyAmorphous1978, fishmanPhaseDiagramsMulticritical1978, wongCompetingOrderParameters1980, 
mukamelPhaseDiagramsMulticritical1981, wongFe1xCoxCl2CompetingAnisotropies1983, defotisMagneticPhaseDiagram1984a, ibarraSingleionCompetingMagnetic1991, katsumataSimultaneousOrderingOrthogonal1992, pirogovTbxEr1xNi5CompoundsIdeal2009, perezPhaseDiagramThreeDimensional2015, bhutaniStrongAnisotropyMixed2020a, foghTuningMagnetoelectricityMixedanisotropy2023}.
Among these, the rare-earth fluoride family $\mathrm{LiRF_4}$, where R denotes a rare-earth ion, has emerged as a particularly clean and tunable platform to explore the effects of long-range dipolar interactions, disorder, anisotropy, and quantum fluctuations \cite{wuClassicalQuantumGlass1991a, 
brookeQuantumAnnealingDisordered1999,
schechterQuantumSpinGlass2006, kraemerDipolarAntiferromagnetismQuantum2012, silevitchFerromagnetContinuouslyTunable2007,liuUltralowfieldMagnetocaloricMaterials2023}.

Within this family, \lhf{} stands out as a paradigmatic Ising dipolar ferromagnet, extensively studied both experimentally and theoretically.
Its diluted variant, \lhfx{}, has served as a test bed for exploring quantum phase transitions, random fields, and glassy dynamics \cite{reichGlassyRelaxationFreezing1987,  wuClassicalQuantumGlass1991a, brookeQuantumAnnealingDisordered1999, giraudNuclearSpinDriven2001, ghoshCoherentSpinOscillations2002, chakrabortyTheoryMagneticPhase2004, schechterSignificanceHyperfineInteractions2005, schechterQuantumSpinGlass2006, tabeiInducedRandomFields2006, biltmoPhaseDiagramDilute2007, schechterLiHoRandomfieldIsing2008, schechterDerivationLowPhase2008, 
tabeiPerturbativeQuantumMonte2008, tamSpinGlassTransitionNonzero2009, gingrasCollectivePhenomenaLiHo2011}.
Recent work revealed the crucial role of off-diagonal dipolar (ODD) terms in these systems, which induce effective internal transverse fields and emergent multispin interactions, leading to substantial corrections to the conventional mean-field description \cite{dollbergEffectIntrinsicQuantum2022b, dollbergLiHoF4SpinhalfNonstandard2024}.

\begin{figure}[h!]
	\centering
	\includegraphics[width=\linewidth]{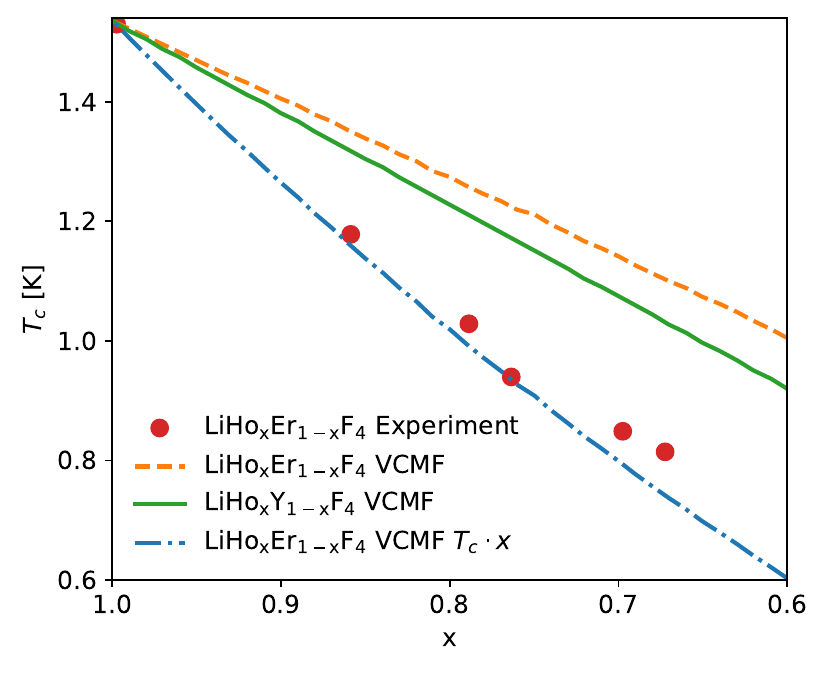}
	\caption{From Piatek \textit{et al.} \cite{piatekPhaseDiagramEnhanced2013}.
		Experimental phase diagram of \lhef{} (red solid circles) compared with virtual crystal mean-field calculations (orange dashed line). The experimental slope is much steeper than the mean-field slope, to the extent that phenomenologically multiplying the VCMF $T_{c}$ by an additional $x$ yields good agreement with the experimental phase diagram (blue dashed-dotted line). As a reference, the mean-field result for \lhfx{} (which is in good agreement with experiment for $x \gtrsim 0.5$) is shown as a green solid line.}
		\label{fig:LiHoErF4-phase-diagram-piatek}
\end{figure}

Compared to \lhf{} and \lhfx{}, \lef{} has received less attention. Its ground state, governed by planar $XY$ anisotropy, results in an antiferromagnetic ordered phase \cite{beauvillainLowtemperatureMagneticSusceptibility1977, kraemerDipolarAntiferromagnetismQuantum2012, liuUltralowfieldMagnetocaloricMaterials2023}.
The mixed-anisotropy system \lhef{}, in which Ising Ho$^{3+}$ ions are randomly replaced by planar Er$^{3+}$ ions, offers a new and largely unexplored playground.
Experimentally, this system was studied by Piatek \textit{et al.} \cite{piatekPhaseDiagramEnhanced2013}, who observed a strikingly rapid suppression of the ferromagnetic transition temperature $T_c$ with Er doping, much steeper than mean-field predictions and even steeper than the linear behavior $T_c(x) = x T_c(x=1)$ observed in \lhfx{} (see Fig~\ref{fig:LiHoErF4-phase-diagram-piatek}).
Moreover, they observed an extended spin-glass phase at intermediate and low concentrations, highlighting the intricate competition between disorder, frustration, and anisotropy.

This surprisingly rapid suppression of $T_c(x)$ calls for a theoretical framework that goes beyond virtual crystal mean-field approximations.
Building on recent insights into the role of ODD terms in \lhf{} and \lhfx{} \cite{dollbergLiHoF4SpinhalfNonstandard2024, dollbergEffectIntrinsicQuantum2022b}, here we demonstrate that the inclusion of such terms, combined with the enhanced susceptibility of planar Er spins to transverse fluctuations, naturally explains the rapid suppression of $T_c$ in \lhef{}.
Our findings shed light on the complex emergent phenomena possible in mixed-anisotropy dipolar magnets and further cement \lhef{} as a uniquely tunable platform for studying disorder-driven and anisotropy-driven phase transitions.

Off-diagonal terms of the dipolar interaction play an essential part in determining the transition temperature of the paramagnetic-ferromagnetic phase transition in \lhf{}.
These terms reduce the energetic cost of fluctuations around the ferromagnetic state by applying internal transverse fields that induce quantum fluctuations, subsequently reducing the energy of the affected spins, leading to a decrease in $T_c$.
Because this effect arises from an emergent three-body interaction \cite{dollbergLiHoF4SpinhalfNonstandard2024}, it is suppressed more rapidly upon dilution---i.e., when Ho ions are randomly substituted with nonmagnetic Y ions to form \lhfx{}---than the dominant two-body longitudinal dipolar interaction.
Consequently, the $T_c(x)$ curve exhibits a milder slope than would be expected in the absence of this mechanism \cite{dollbergEffectIntrinsicQuantum2022b}.

In this paper, we argue that a similar mechanism is at work in \lhef{}, but that it is enhanced rather than weakened upon dilution, due to the mixed Ising-$XY$ anisotropy that amplifies the effect of ODD interactions.
Thus, proper inclusion of off-diagonal dipolar terms and analysis beyond mean-field are required to understand deviations from theoretical predictions in \lhef{} and correctly reproduce the experimental $T_{c}\left(x\right)$ curve.

\section[Virtual Crystal Mean-Field Theory of $\mathrm{LiHo_{x}Er_{1-x}F_{4}}$]{Virtual Crystal Mean-Field Theory of $\mathbf{LiHo_{x}Er_{1-x}F_{4}}$} \label{sec:VCMF}

The mean-field calculation whose results are shown in Fig.~\ref{fig:LiHoErF4-phase-diagram-piatek} is known as the virtual crystal mean field (VCMF) \cite{dallapiazzaMeanFieldCalculationsDiluted2009,piatekUltraLowTemperature2012,kraemerDipolarAntiferromagnetismQuantum2012}.
Within this framework, two mean-field Hamiltonians are considered, corresponding to each ion type, and the mean-field is taken to be a linear combination of the two self-consistent fields in proportions dictated by the concentration $x$. The VCMF Hamiltonian corresponding to ion species $t$ situated at sublattice $k$ can be expressed as:
\begin{multline}
	\mathcal{H}_{k,t}^{\mathrm{VCMF}}=\boldsymbol{J}_{k,t}\cdot \left(x\boldsymbol{H}_{k,\mathrm{Ho}}+\left(1-x\right)\boldsymbol{H}_{k,\mathrm{Er}}\right)\\+V_{\mathrm{C}}^{t}(\boldsymbol{J}_{k,t})+A^{t}\boldsymbol{J}_{k,t}\cdot\boldsymbol{I}_{k,t}.\label{eq:VCMF-Hamiltonian}
\end{multline}
Here, the index $t$ explicitly denotes the ion type, $t\in\{\mathrm{Ho},\mathrm{Er}\}$, and all quantities carrying this index---such as
$V^{t}_{C}$, $A^{t}$, and $\mathbf{H}_{k,t}$---take the values corresponding to that specific rare-earth species.

$V_{\mathrm{C}}^{t}$ is the crystal-field Hamiltonian; $\boldsymbol{J}_{k,t}$ and $\boldsymbol{I}_{k,t}$ are, respectively, the electronic and nuclear angular momentum vector operators for either Ho or Er; and $\boldsymbol{H}_{k}^{\mathrm{Ho}}$, $\boldsymbol{H}_{k}^{\mathrm{Er}}$ are the self-consistent fields induced by $\mathcal{H}_{k,\mathrm{Ho}}^{\mathrm{VCMF}}$ and $\mathcal{H}_{k,\mathrm{Er}}^{\mathrm{VCMF}}$, respectively. $A^{t}$ denotes the hyperfine interaction strength; its values are given in Appendix~\ref{app:eff-H-LiHoErF4}.
Given the antiferromagnetic nature of the ordered phase in \lef{}, it is necessary to consider sublattices, as indicated by the index $k$.
We consider $k=1,\ldots,4$ for the four rare-earth ions per unit cell \cite{kraemerDipolarAntiferromagnetismQuantum2012}, with the dipolar interactions between the sublattices calculated by the Ewald method \cite{wangEstimateCutoffErrors2001}. Off-diagonal interactions are eliminated due to symmetry within the VCMF approximation.

\begin{figure}[ht]
	\centering
	\includegraphics[width=\linewidth]{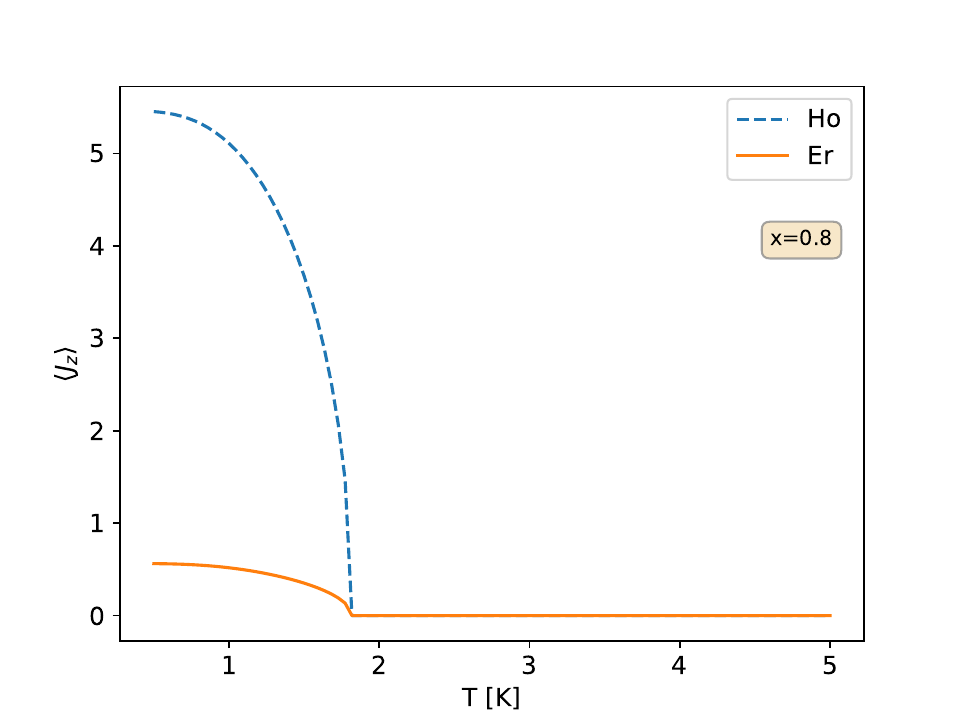}
	\caption{The expectation value $\langle{J^z}\rangle$ of the Ho and Er ions vs. temperature in $\mathrm{LiHo_{0.8}Er_{0.2}F_{4}}$ within the VCMF approximation. Despite the planar XY anisotropy induced by the crystal-field Hamiltonian of the Er ions, nonzero $\langle{J^{z}}\rangle$ values are found below the critical temperature.}
	\label{fig:Jz-vs-T-LiHoErF4}
\end{figure}

The erbium ions exhibit a relatively mild anisotropy ratio, with $g_{\perp}/g_{\parallel} \approx 3 $ \cite{hansenMagneticPropertiesLithium1975, magarinoEPRExperimentsLiTb1980, kraemerDipolarAntiferromagnetismQuantum2012}. Although they preferentially align their magnetic moment in the plane perpendicular to the holmium's Ising axis, they can moderately polarize along the perpendicular axis as well. This behavior offers insight into why the VCMF approximation yields a milder slope for \lhef{} compared to \lhfx{}, a prediction that stands in contrast to experimental observations.
As Fig.~\ref{fig:Jz-vs-T-LiHoErF4} shows, within the VCMF approach, the Er ion indeed develops a small magnetic moment in the $z$ direction in the ferromagnetic phase, despite its pronounced planar XY anisotropy. This behavior contrasts with Y ions, which are nonmagnetic and manifest virtually no magnetic moment in any direction. The milder slope in \lhef{} can therefore be attributed to the fact that the $z$ component of the mean-field acting on the Ho ions is not merely reduced in proportion to $x$, but rather mixed with the weaker (yet still nonzero) field generated by the Er ions, as detailed in Eq.~\eqref{eq:VCMF-Hamiltonian}.
That said, the larger question remains: why is the experimental $T_{c}\left(x\right)$ slope so much steeper than the VCMF one, steeper even than that of \lhfx{}?
To answer this question, we must consider ODD interactions between the Ho and Er spins, neglected in the VCMF due to symmetry.

\section[Off-diagonal dipolar interactions in $\mathrm{LiHo_{x}Er_{1-x}F_{4}}$]{Off-diagonal dipolar interactions in $\mathbf{LiHo_{x}Er_{1-x}F_{4}}$}
The long-range dipolar interaction between two magnetic moments located at lattice sites $i$ and $j$ is given by
\begin{equation}
V_{ij}^{\mu\nu}=\frac{\delta_{\mu\nu}r_{ij}^2-3\,r_{ij}^{\mu}r_{ij}^{\nu}}{r_{ij}^5},
\label{eq:dipolar-int}
\end{equation}
where $\mathbf{r}_{ij}$ is the displacement vector between the two ions and $\mu,\nu \in \{x,y,z\}$ denote Cartesian components. In \lhf{}, the diagonal longitudinal components ($\mu=\nu=z$) induce ferromagnetic order below $T_c$, while the off-diagonal components ($\mu\!\ne\!\nu$)---particularly $V_{ij}^{xz}$ and $V_{ij}^{yz}$---were shown to reduce said $T_c$ by generating virtual excitations between crystal-field levels in what can be expressed as an effective three-body interaction \cite{dollbergEffectIntrinsicQuantum2022b, dollbergLiHoF4SpinhalfNonstandard2024}.

As with \lhfx{}, the slope of the $T_c\left( x \right)$ curve arises from contributions due to (i) the reduction of the dominant longitudinal dipolar interactions with dilution, and (ii) the modulation---either attenuation or enhancement---of effective interactions induced by off-diagonal dipolar terms.
As explained in Ref.~\cite{dollbergEffectIntrinsicQuantum2022b}, in the case of \lhfx{}, the off-diagonal dipolar mechanism diminishes as Ho is diluted by Y.

\begin{figure}[ht]
	\centering
	\includegraphics[width=\linewidth]{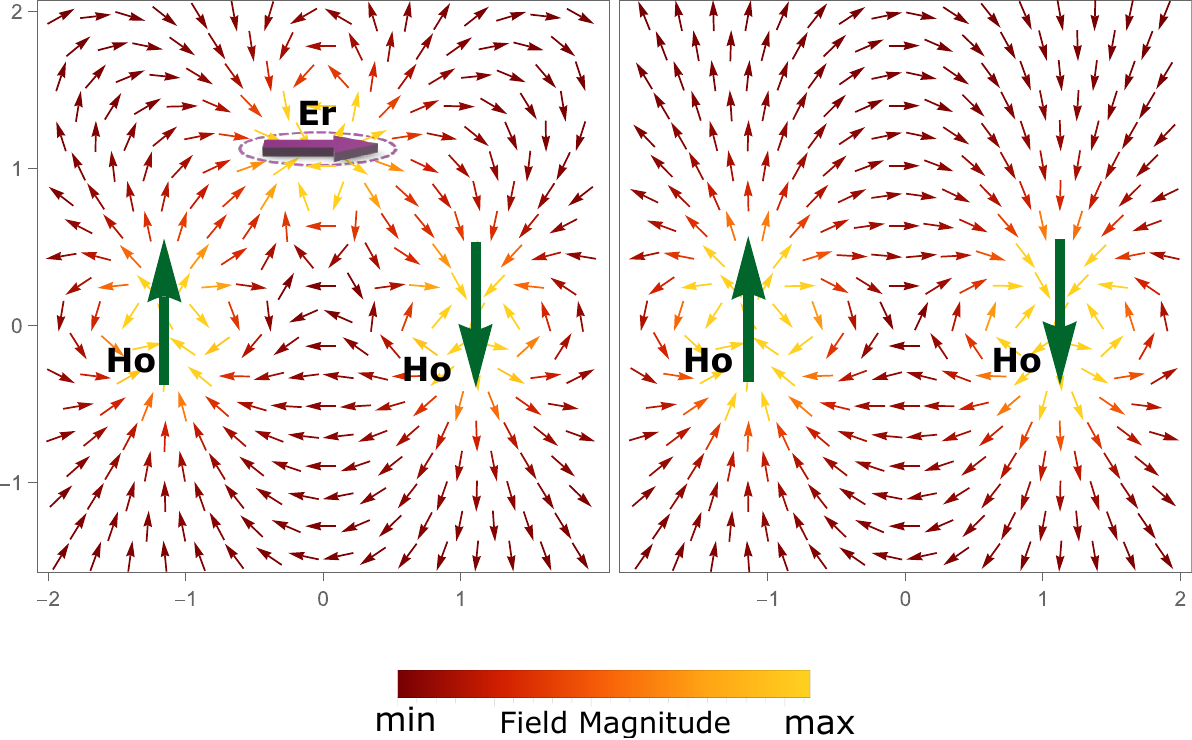}
	\caption{A sketch of three ions and their combined
		generated magnetic field. On the left, the middle ion is an Er ion
		(purple arrow), oriented with the transverse field exerted on it by the
		two Ho ions (green arrows). On the right, the middle ion is missing,
		signifying a nonmagnetic Y ion. The energetic contribution of off-diagonal
		dipolar terms is zero for the configuration on the right and negative
		for the configuration on the left.}
		\label{fig:mechanism-sketch-Er}
\end{figure}

In contrast, we expect an analogous mechanism in \lhef{} to be enhanced by the modest substitution of Ho with Er. Because the magnetic moments of the Er ions lie in the plane perpendicular to the Ising axis of the Ho ions, they exert---via their ODD interactions---a longitudinal magnetic field that energetically favors antiferromagnetic correlations between surrounding Ho ions. This scenario is illustrated in Fig.~\ref{fig:mechanism-sketch-Er}. Such antiferromagnetic correlations, strengthened by the presence of intermediate Er ions, ultimately lead to a more pronounced suppression of the ferromagnetic transition temperature. Specifically, when the central ion in Fig.~\ref{fig:mechanism-sketch-Er} is an Er ion, the effective antiferromagnetic interaction between the two Ho ions is enhanced compared to when the central ion is nonmagnetic (Y).
To further explore and quantify this mechanism, we next turn our attention to the particulars of our model in the following section.

\section{Details of the model}
\label{sec:details-of-model}
In this section we first present the full microscopic Hamiltonian, then motivate a low-energy classical approximation appropriate for studying the concentration dependence of $T_c$, and finally define the effective Hamiltonian used in the Monte Carlo simulations.

The full microscopic Hamiltonian of \lhef{} is given by
\begin{equation}\label{eq:micro-hamiltonian-LiHoErF4}
	\mathcal{H} = \mathcal{H}^{\mathrm{Ho}} + \mathcal{H}^{\mathrm{Er}} + \mathcal{H}^{\mathrm{Ho-Er}},
\end{equation}
where each term represents interactions among or between the rare-earth ions occupying the four sublattices of the tetragonal scheelite crystal structure.

For each ion type $t\!\in\!\{\mathrm{Ho},\mathrm{Er}\}$,\begin{multline}\label{eq:micro-hamiltonian-ion-specific}
	\mathcal{H}^{t} = \sum_i \epsilon_i^t V_C^{t}(\boldsymbol{J}_{i,t}) 
    + \frac{1}{2}
	(g_L^t \mu_B)^2 \frac{\mu_0}{4\pi} \sum_{\substack{i\neq j\\\mu,\nu}} \epsilon_i^t \epsilon_j^t V_{ij}^{\mu \nu} J_{i,t}^{\mu} J_{j,t}^{\nu} \\
	+ J_{\text{ex}}^t
	\sum_{\left\langle i,j \right\rangle} \epsilon_i^t \epsilon_j^t \boldsymbol{J}_{i,t} \cdot
	\boldsymbol{J}_{j,t}
	+ A^t \sum_i \epsilon_i^t (\boldsymbol{I}_{i,t} \cdot \boldsymbol{J}_{i,t}).
\end{multline}
The variables $\epsilon_i^t = \{0,1\}$ denote the occupation of an ion of type $t$ at site $i$ on the lattice (so we cannot have $\epsilon_i^{\mathrm{Ho}}=1$ and $\epsilon_i^{\mathrm{Er}}=1$ for the same $i$).
The script $t$ indicates to which ion type the expression refers, with numerical values for the Land{\'e} g-factors $g_L^{t}$, the nearest-neighbor exchange interactions $J_{\text{ex}}^t$, and the hyperfine interaction strengths $A^t$ given in Appendix~\ref{app:eff-H-LiHoErF4}.
The $V_C^{t}(\boldsymbol{J}_{i,t})$ term is an anisotropic crystal-field potential that imposes either an Ising easy axis (for holmium), or an $XY$ easy plane (for erbium).
The mutual interaction between the two ion types is given by
\begin{equation}\label{eq:micro-hamiltonian-diff-types}
	\mathcal{H}^{\mathrm{Ho-Er}} = \dfrac{\mu_0}{4 \pi} \mu_B^2 \sum_{i \neq j} \sum_{\mu,\nu} g_L^{\mathrm{Ho}} g_L^{\mathrm{Er}} \epsilon_i^{\mathrm{Ho}} \epsilon_j^{\mathrm{Er}} V_{ij}^{\mu \nu} J_{i,\mathrm{Ho}}^{\mu} J_{j,\mathrm{Er}}^{\nu}.
\end{equation}

When treating the full microscopic Hamiltonian~\eqref{eq:micro-hamiltonian-LiHoErF4}, our objective is twofold. First, we aim to incorporate the quantum effects established in previous works, specifically those related to the interactions among Ho ions and the corrections arising from the coupling of Ising states to an excited state at $\Delta\approx \qty{10}{\kelvin}$. Second, we wish to capture the essentially classical effect of ODD interactions between Ho and Er ions. The former has a very modest impact on the $T_c(x)$ slope (see Fig.~3 in Ref.~\cite{dollbergEffectIntrinsicQuantum2022b}) compared to the discrepancy between experiment and VCMF observed in Fig.~\ref{fig:LiHoErF4-phase-diagram-piatek}.

Hence, we argue that a low-energy classical approximation that accounts only for pairwise dipolar and exchange interactions should suffice to capture the \emph{concentration dependence} of the critical temperature $T_c(x)$.
In contrast, reproducing the \emph{absolute value} of $T_c$ at full holmium concentration, $x = 1$, requires the inclusion of quantum-derived emergent interactions that would not appear in a naive low-energy projection. As demonstrated in Refs.~\cite{dollbergEffectIntrinsicQuantum2022b,dollbergLiHoF4SpinhalfNonstandard2024}, these emergent interactions take the form of three-body interactions, and result in a notable suppression of $T_c$ at $x = 1$, while having relatively little influence on the slope of $T_c(x)$ at concentrations $x<1$.
We accordingly proceed by projecting the full Hamiltonian~\eqref{eq:micro-hamiltonian-LiHoErF4} onto a classical model, followed by the explicit inclusion of the emergent interaction induced by quantum crystal-field excitations. 

In practice, this approximation is realized by substituting the angular momentum operators with effective classical variables. For \Er{} ions, which fluctuate within the $xy$-plane due to their planar anisotropy, we take $\hat{J}^{x}_{i, \mathrm{Er}} \rightarrow J_{xy} \cos\theta_i$ and $\hat{J}^{y}_{i, \mathrm{Er}} \rightarrow J_{xy} \sin\theta_i$, with $\theta_i \in [0, 2\pi)$. For \Ho{} ions, which are Ising-like, we replace $\hat{J}^{z}_{i, \mathrm{Ho}} \rightarrow \alpha s_i$ with $s_i \in {\pm 1}$. The values of $J_{xy}$ and $\alpha$, which characterize the effective moment magnitudes consistent with the crystal-field-induced anisotropies, are specified in Appendix~\ref{app:eff-H-LiHoErF4}.
Within this essentially zeroth-order approximation, operators that do not conform to the anisotropy imposed by the respective crystal-field potential are neglected. This leads us to set $\hat{J}^{z}_{i, \mathrm{Er}} \to 0$ and discard transverse \Ho{} operators, i.e., $\hat{J}^{x,y}_{i, \mathrm{Ho}} \to 0$, except to the extent that they contribute indirectly through the quantum-derived three-body interactions mentioned above.
In addition, we neglect the hyperfine interaction entirely, as it only becomes significant at low temperatures \cite{bitkoQuantumCriticalBehavior1996, schechterSignificanceHyperfineInteractions2005, schechterDerivationLowPhase2008}.

The resulting Hamiltonian, utilized for the simulation, is given by
\begin{align}\label{eq:MC-hamiltonian-LiHoErF4}
	\mathcal{H}_{\mathrm{MC}} &= 
	\frac{1}{2}
	(g_L^{\mathrm{Ho}} \mu_B \alpha)^2 \frac{\mu_0}{4\pi} \sum_{i\neq j} \epsilon_i^{\mathrm{Ho}} \epsilon_j^{\mathrm{Ho}} V_{ij}^{zz} s_{i}^{z} s_{j}^{z} \nonumber\\
    &+ \Jex \alpha^2
	\sum_{\left\langle i,j \right\rangle} \epsilon_i^{\mathrm{Ho}} \epsilon_j^{\mathrm{Ho}} s_{i}^{z} s_{j}^{z} \nonumber\\
    &+ \frac{1}{2}
	(g_L^{\mathrm{Er}} \mu_B)^2 \frac{\mu_0}{4\pi} \sum_{i\neq j} \sum_{(\mu, \nu) = (x,y)}
	\epsilon_i^{\mathrm{Er}} \epsilon_j^{\mathrm{Er}}
	V_{ij}^{\mu \nu} \left( \vec{\tau}_{i} \right)^{\mu} \left( \vec{\tau}_{j} \right)^{\nu}\nonumber\\
	&+\alpha\dfrac{\mu_0}{4 \pi} \mu_B^2 \sum_{i \neq j} \sum_{\mu \in \left\{x,y\right\}} g_L^{\mathrm{Ho}} g_L^{\mathrm{Er}} \epsilon_i^{\mathrm{Ho}} \epsilon_j^{\mathrm{Er}} V_{ij}^{\mu z} s_{i}^{z} \left( \vec{\tau}_{j} \right)^{\mu}\nonumber\\
	&-K x \alpha^2 \sum_{i\ne j} \epsilon_i^{\mathrm{Ho}} \epsilon_j^{\mathrm{Ho}} \sum_{k}\left(V_{ik}^{xz}V_{kj}^{xz}+V_{ik}^{yz}V_{kj}^{yz}\right)s_{i}^{z}s_{j}^{z} ,
\end{align}
where $s^z = \pm 1$ and $\vec{\tau} = J_{xy} \left(\cos \theta, \sin \theta  \right)$ are variables set and varied by the MC framework.
The first two terms, respectively, denote the dipolar and exchange Ho-Ho interactions.
The third and fourth terms describe the dipolar Er-Er and Ho-Er interactions, respectively, and the last term describes the interactions between Ho ions induced by quantum crystal-field excitations; its form and magnitude are explained in Appendix~\ref{app:eff-H-LiHoErF4}.
All sums over $i$, $j$, and $k$ run over all rare-earth sites, except in the second term, where the sum is over all distinct nearest-neighbor pairs.

In addition to \lhef{}, we also simulate the diluted Ising magnet \lhfx{} within the same Monte Carlo framework.
Operationally, this is implemented by retaining only the holmium degrees of freedom with site occupation
$\epsilon_i^{\mathrm{Ho}}=1$ occurring with probability $x$, and removing all erbium degrees of freedom from the Hamiltonian \eqref{eq:MC-hamiltonian-LiHoErF4}.
Equivalently, this may be viewed as taking $\epsilon_i^{\mathrm{Er}}=0$ on all sites, so that only Ho-Ho interactions remain.
The resulting model thus consists of the Ho-Ho dipolar, exchange, and  effective interactions described respectively by the first, second, and last terms in $\mathcal{H}_{\mathrm{MC}}$ above.

This construction highlights an important limiting case of Eq.~\eqref{eq:MC-hamiltonian-LiHoErF4}.
Within the present approximation, Ho ions carry only Ising ($z$) components, while Er ions carry only planar ($x$--$y$) components.
As a result, the coupling between Ho and Er degrees of freedom arises exclusively from the off-diagonal dipolar term
proportional to $V_{ij}^{\mu z}$ (with $\mu \in \left\{x,y\right\}$) in the fourth line of Eq.~\eqref{eq:MC-hamiltonian-LiHoErF4}.
When this term is absent---either because Er ions are removed, as in \lhfx{}, or because the Ho-Er off-diagonal dipolar interaction
is explicitly set to zero---the Ho and Er sectors decouple completely.
In this decoupled limit, the ferromagnetic transition of the Ho subsystem is independent of the Er configuration and coincides
with that of \lhfx{} at the same Ho concentration.

Consequently, within the present model, any difference between the Monte Carlo $T_c(x)$ curves of \lhef{} and \lhfx{}
directly isolates the effect of Ho-Er off-diagonal dipolar interactions.

\section[Monte Carlo simulation results]{Monte Carlo simulation results}

\begin{figure}[h]
	\centering
	\includegraphics[width=\linewidth]{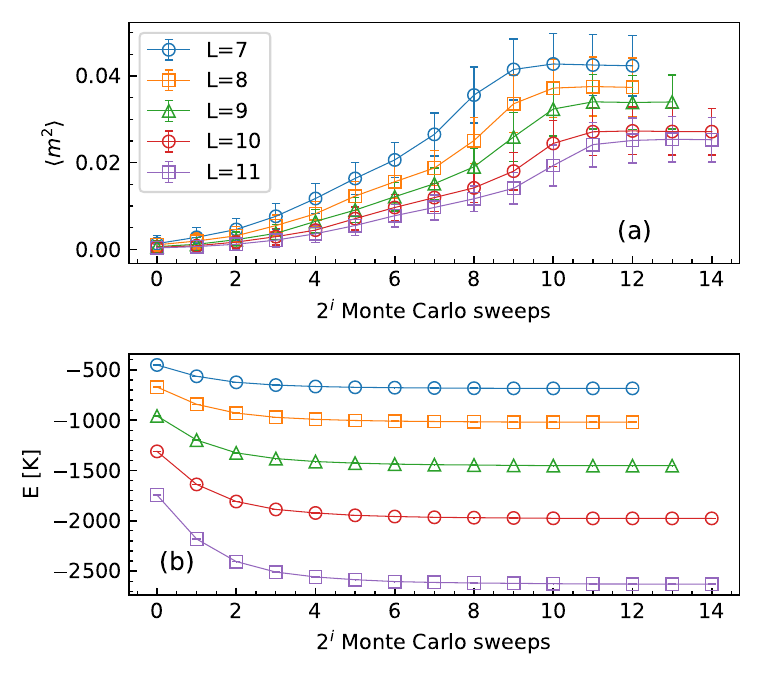}
	\caption{Equilibration of $\mathrm{LiHo_{0.675}Er_{0.325}F_{4}}$ at $T \approx \qty{0.87}{\kelvin}$, the simulated temperature closest to $T_c$.
(a) Second moment of the magnetization, $\langle m^2 \rangle$, and (b) average energy $E$, as functions of logarithmically binned Monte Carlo sweeps.
Each point represents an average over binned data, where the $i$-th bin includes $2^{i}$ Monte Carlo sweeps (e.g., 1, 2--3, 4--7, etc.).
Equilibrium is assumed when the final three points agree within their statistical error bars.}
	\label{fig:equilibration}
\end{figure}

\begin{figure}[h]
	\centering
	\includegraphics[width=\linewidth]{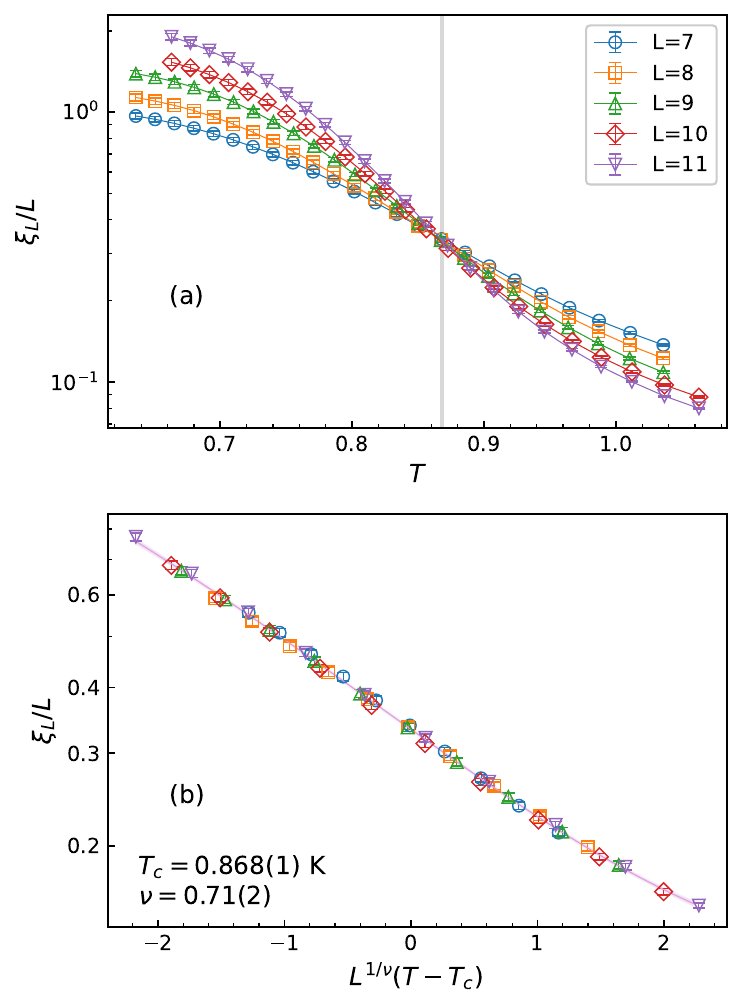}
	\caption{Finite-size scaling analysis of Monte Carlo simulation data for $\mathrm{LiHo_{0.675}Er_{0.325}F_{4}}$.
(a) Finite-size correlation length divided by the linear system size, $\xi_L/L$, as a function of temperature $T$ for system sizes $L = 7$, $8$, $9$, $10$, and $11$. The curves evidently cross at the critical temperature $T_c = \qty{0.868(1)}{\kelvin}$, indicated by the shaded region, whose value and uncertainty are determined from the finite-size scaling analysis in panel (b).
(b) Scaling collapse of $\xi_L/L$ according to the finite-size scaling form, using $T_c = \qty{0.868(1)}{\kelvin}$ and critical exponent $\nu = 0.71(2)$. The solid line represents the scaling function inferred by BSA demonstrating that, with these parameters, data for different system sizes collapse onto a single universal curve.}
	\label{fig:finite-size-scaling}
\end{figure}

We study the classical effective Hamiltonian~\eqref{eq:MC-hamiltonian-LiHoErF4} using
parallel-tempering (exchange) Monte Carlo~\cite{hukushimaExchangeMonteCarlo1996},
combined with single-site Metropolis updates~\cite{newmanMonteCarloMethods1999, bergMarkovChainMonte2004}.
We simulate boxes of linear size $L$ (containing $N=4L^3$ rare-earth sites) with periodic boundary conditions. All long-range dipolar sums are evaluated using the Ewald
method, yielding effective pair couplings appropriate to the periodically replicated system \cite{wangEstimateCutoffErrors2001}. We do not include a demagnetization term, consistent with the needle-shaped-domain picture commonly adopted for \lhf{} magnets \cite{chakrabortyTheoryMagneticPhase2004,tabeiPerturbativeQuantumMonte2008,tamSpinGlassTransitionNonzero2009, mckenzieFluctuationsPhaseTransitions2016}.

For each concentration $x$, we generate independent quenched disorder realizations by
assigning $\epsilon_i^{\mathrm{Ho}}=1$ with probability $x$ and $\epsilon_i^{\mathrm{Er}}=1$ with probability
$1-x$ (with no double occupancy). For each realization we simulate $N_T$ replicas at different
temperatures $\{T_n\}$ spanning $[T_{\min},T_{\max}]$ (Table~\ref{table:sim-params-LiHoErF4}).
A Monte Carlo sweep (MCS) consists of $N$ attempted single-site updates plus a full set of
replica-exchange attempts between neighboring temperatures.
For a \Ho{} site, we propose an Ising flip $s_i^z\to -s_i^z$; for an \Er{} site, we propose a
rotation $\theta_i\to \theta_i'\in[0,2\pi)$ with $\theta_i'$ drawn uniformly, and accept/reject by the Metropolis criterion based on the energy change computed
from~\eqref{eq:MC-hamiltonian-LiHoErF4}.
Replica exchanges between adjacent temperatures $T_n$ and $T_{n+1}$ are attempted with the
standard parallel-tempering acceptance probability,
$\min\{1,\exp[(\beta_n-\beta_{n+1})(E_{n+1}-E_n)]\}$, which substantially improves sampling
near criticality and in the presence of disorder.

For each parameter set $(x,L)$, we run $2^b$ MCS for equilibration followed by an additional
$2^b$ MCS for measurements (see Table~\ref{table:sim-params-LiHoErF4} for the values of $b$
and the number of disorder realizations $N_{sa}$).
Equilibration is verified using the logarithmic-binning procedure: Monte Carlo time series are grouped into bins whose lengths grow as $2^i$, and observables are
averaged within each bin. The simulation is deemed equilibrated once all observables of
interest (in practice, the energy and the moments entering $\xi_L$) agree within statistical
errors for three consecutive bins \cite{katzgraberUniversalityThreedimensionalIsing2006}.
An example of this criterion applied close to criticality is shown in Fig.~\ref{fig:equilibration}, where both $\langle m^2\rangle$ and $E$ plateau within errors in the final bins.

Thermal averages $\langle\cdots\rangle_T$ are computed from the post-equilibration segment of
each replica trajectory; quenched averages $[\cdots]_{\rm av}$ are then obtained by averaging
over the $N_{sa}$ disorder realizations (Table~\ref{table:sim-params-LiHoErF4}).

To locate the ferromagnetic transition we compute the finite-size correlation length
$\xi_L$ using the wavevector
$\mathbf{k}_{\min}=(2\pi/L,0,0)$ \cite{ballesterosCriticalBehaviorThreedimensional2000a}:
\begin{equation}
\xi_L=\frac{1}{2\sin(k_{\min}/2)}
\left(
\frac{\big[\langle m^2(\mathbf{0})\rangle_T\big]_{\rm av}}
     {\big[\langle m^2(\mathbf{k}_{\min})\rangle_T\big]_{\rm av}}-1
\right)^{1/2},
\end{equation}
with
$m(\mathbf{k})=\frac{1}{N}\sum_{i} s^z_i\,e^{-i\mathbf{k}\cdot\mathbf{R}_i}$.
The dimensionless ratio $\xi_L/L$ obeys the scaling form
$\xi_L/L=\tilde{X}\!\left(L^{1/\nu}(T-T_c)\right)$ near the transition, implying crossings of $\xi_L/L$ for different $L$ at $T_c$.

\begin{figure}[h!]
    \centering
    \includegraphics[width=\linewidth]{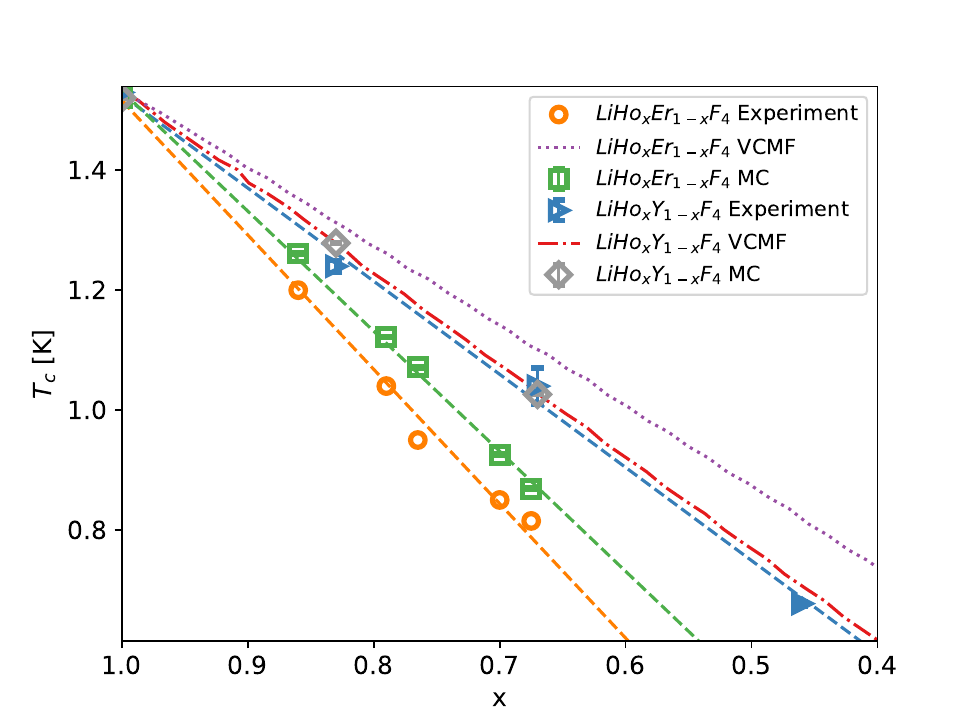}
    \caption{$T-x$ phase diagram of \lhef{} and \lhfx{}. Experimental results from Refs.~\cite{babkevichPhaseDiagramDiluted2016,piatekPhaseDiagramEnhanced2013} are denoted by triangles for \lhfx{} and circles for \lhef{}.
    VCMF results for \lhfx{} (\lhef{}), given by the dot-dashed (dotted) line, are reproduced in this work based on Refs.~\cite{piatekCharacterisationLiHoxEr1XF42009,piatekUltraLowTemperature2012,piatekPhaseDiagramEnhanced2013}.
    Monte Carlo results for \lhef{} (\lhfx{}) are denoted by squares (diamonds).
    Some results are accompanied by linear best-fit lines as guides for the eye.}
    \label{fig:LiHoErF4-x-T-phase-diagram}
\end{figure}

We extract $T_c$ and $\nu$ by Bayesian scaling analysis (BSA)~\cite{haradaBayesianInferenceScaling2011a,haradaKernelMethodCorrections2015}, using the reference C++ code provided with the method~\cite{haradaKernelMethodCorrections2015}. BSA uses Gaussian-process regression to infer the finite-size scaling function without assuming a polynomial form. Parameter estimates and confidence intervals are obtained using a hybrid Monte Carlo sampling procedure implemented in the software~\cite{haradaKernelMethodCorrections2015}.
The resulting crossing and collapse for a representative concentration are shown in Fig.~\ref{fig:finite-size-scaling}, and the corresponding $T_c(x)$ data points are reported in
Fig.~\ref{fig:LiHoErF4-x-T-phase-diagram} and Table~\ref{table:sim-params-LiHoErF4}.

\section{Discussion and Conclusion}

A prompt comparison reveals that, compared to the VCMF approach, a MC simulation incorporating ODD interactions between Ho and Er spins shows a significantly steeper $T_c(x)$ curve. In fact, the MC curve has a larger absolute slope compared to both the VCMF for \lhef{} and the VCMF for \lhfx{} (the latter is in good agreement with the respective experimental curve within the dilution range currently under consideration), putting it decidedly closest to the experimental curve.

Notably, the rapid suppression of $T_c(x)$ in \lhef{}, which VCMF fails to capture, cannot be attributed solely to generic magnetic frustration associated with Er substitution.
For the concentration range studied here, the system retains a robust ferromagnetic ground state with simple long-range order, and Er moments do not introduce competing macroscopic orders.
Instead, the ODD Ho-Er couplings generate specific local correlations that lower the energy of certain Ho configurations while preserving global ferromagnetic order.
Virtual crystal mean-field theory, which averages over such local correlations, is therefore insensitive to this mechanism.
The resulting discrepancy between VCMF predictions and experiment highlights the importance of fluctuations and correlations beyond mean field that are not associated with extensive degeneracy or loss of order, but rather with the tensorial nature of dipolar interactions in mixed-anisotropy systems.

Beyond comparison to VCMF, an especially informative benchmark is the Monte Carlo result for \lhfx{}. Within the present approximation (strictly planar Er and strictly Ising Ho), \lhfx{} is not merely a different material but serves as a control model for \lhef{} with Ho-Er off-diagonal dipolar couplings removed: setting the Ho-Er ODD term to zero decouples the Ho and Er sectors, leaving the Ho transition governed solely by the Ho-Ho interactions and therefore identical to the diluted Ho-Y model. The steeper suppression of $T_c(x)$ that we obtain for \lhef{} relative to \lhfx{} thus isolates a single physical origin---Ho-Er off-diagonal dipolar interactions---and provides a direct numerical demonstration that these couplings are sufficient to generate an enhanced reduction of the Ho ferromagnetic transition temperature.

Nevertheless, certain limitations of the current model warrant acknowledgment. For instance, the magnitudes of the Er magnetic moments were kept constant during simulations; a more comprehensive treatment would allow these magnitudes to vary with the local field (see Fig.~\ref{fig:Er-Jx-vs-Bx}). Additionally, Er ions were modeled as purely planar spins, whereas, in fact, they might exhibit a slight out-of-plane tilt. Accounting for this tilt would likely diminish the observed effect to some extent, and it is thus an important consideration for a comprehensive understanding that goes beyond our current results.

The apparent increase of the extracted values of $\nu$ with dilution, as obtained from our finite-size scaling analysis (Table~\ref{table:sim-params-LiHoErF4}), suggests a crossover away from the pure \lhf{} critical behavior upon dilution of Ho by either Er or Y, consistent with theoretical prediction \cite{harrisEffectRandomDefects1974} and numerical evidence from other disordered 3D Ising systems \cite{ballesterosCriticalExponentsThreedimensional1998, calabreseThreedimensionalRandomlyDilute2003, bercheBondDilution3D2004, hasenbuschUniversalityClass3D2007, hasenbuschCriticalBehaviorThreedimensional2007}. However, further studies using larger system sizes and a more systematic treatment of corrections to scaling will be needed to determine the asymptotic universality class of \lhef{}.

The mixed-anisotropy Ising--$XY$ magnet \lhef{} provides an opportunity to explore a unique facet of the role ODD terms play in different members of the $\mathrm{LiREF_4}$ family.
Specifically, it presents the opportunity to study a classical version of the ODD-induced mechanism described in recent works. Here, ODD-induced internal fields promote antiferromagnetic Ho configurations, making them more prevalent and thus leading to a stronger suppression of $T_c$, without involving quantum fluctuations.

This investigation is motivated by the experimental findings of Piatek \textit{et al.}, who reported a notably rapid decrease of the paramagnetic--ferromagnetic transition temperature with decreasing Ho concentration. 
To address the discrepancy between these observations and VCMF predictions, we undertake a detailed theoretical examination.
First, we show that the shallower slope of $T_c(x)$ obtained through the VCMF approach for \lhef{}, compared to the naive form $T_c(x)=x T_c(x=1)$ observed for \lhfx{}, can be explained by the finite longitudinal magnetic polarization of Er ions.
Furthermore, we argue that the experimentally observed slope, steeper than even the naive dilution prediction, arises from the distinct role of ODD terms in the two systems. Specifically, while dilution tends to mitigate the influence of ODD terms in \lhfx{}, it amplifies their impact in \lhef{}, resulting in a further decrease in $T_c$.
To quantitatively assess this mechanism, we develop a computational model that incorporates detailed properties of Ho spins based on established frameworks and represents Er spins within a simplified classical $XY$ approximation. 
Numerical simulations support our hypothesis by reproducing a steeper $T_c(x)$ curve, underscoring the critical role of ODD interactions between Ho and Er ions.
In conclusion, our work clarifies the intricate interplay of anisotropy, disorder, and ODD interactions in \lhef{}, highlighting the necessity of considering these interactions to accurately describe the physics of mixed-anisotropy dipolar magnetic systems.

\begin{acknowledgments}
M.S. acknowledges support from the Israel Science Foundation (Grant No. 3679/24).
\end{acknowledgments}

\section*{Data Availability}
The data that support the findings of this article are not publicly available upon publication because it is not technically feasible and/or the cost of preparing, depositing, and hosting the data would be prohibitive within the terms of this research project. The data are available from the authors upon reasonable request.

\appendix
\section{Effective Hamiltonian for $\mathbf{LiHo_{x}Er_{1-x}F_{4}}$} \label{app:eff-H-LiHoErF4}
In this appendix, we elaborate on the Hamiltonian \eqref{eq:MC-hamiltonian-LiHoErF4}, used to study the \lhef{} system via MC simulations. The microscopic Hamiltonian \eqref{eq:micro-hamiltonian-LiHoErF4} introduces a few parameters, decorated by a script $t$, whose values differ between Ho and Er ions.
So, $g_L^{t}$ is the Land{\'e} g-factor of either holmium or erbium, where $g_L^{\mathrm{Ho}}=1.25$ and $g_L^{\mathrm{Er}}=1.2$. Further, $V_C^{t}$ is the crystal-field Hamiltonian of either holmium or erbium \cite{ronnowMagneticExcitationsQuantum2007, kraemerDipolarAntiferromagnetismQuantum2012}, and $\boldsymbol{J}_{i,t}$ are the angular momentum operators of the holmium and erbium electrons, which have $J=8$ and $J=15/2$, respectively.
The last term in Eq.~\eqref{eq:micro-hamiltonian-ion-specific} describes the hyperfine interaction between the Ho or Er electrons and their host ion's nucleus. For this interaction, we have established values of $A^{\mathrm{Ho}} = \qty{0.039}{\kelvin}$ and $A^{\mathrm{Er}} = \qty{0.005}{\kelvin}$, but should also note that only \qty{23}{\percent} of Er nuclei possess spins, as cited in \cite{kraemerDipolarAntiferromagnetismQuantum2012}.
Having said that, our findings indicate that the impact of the hyperfine interaction on the \lhef{} VCMF phase diagram is entirely negligible, consistent with earlier studies of \lhf{} and \lef{} that reported a meaningful effect only at much lower temperatures \cite{bitkoQuantumCriticalBehavior1996,schechterSignificanceHyperfineInteractions2005, schechterDerivationLowPhase2008, kraemerDipolarAntiferromagnetismQuantum2012}. Therefore, consistent also with Refs.~\cite{dollbergEffectIntrinsicQuantum2022b,dollbergLiHoF4SpinhalfNonstandard2024}, we continue to neglect the hyperfine interaction in the derivation of $\mathcal{H}_{\mathrm{MC}}$ \eqref{eq:MC-hamiltonian-LiHoErF4}, setting $A^{\mathrm{Ho}}=A^{\mathrm{Er}}=0$.
The exchange interaction also differs between the two ion types, where $J_{\text{ex}}^{\mathrm{Ho}} = \qty{1.16}{\milli\kelvin}$ and $J_{\text{ex}}^{\mathrm{Er}} = 0$ \cite{kraemerDipolarAntiferromagnetismQuantum2012}. This absence of exchange interaction for the Er ion is the reason there are only Ho-Ho exchange interactions in $\mathcal{H}_{\mathrm{MC}}$ \eqref{eq:MC-hamiltonian-LiHoErF4}.

For the classical description adopted in the derivation, we set the magnitude of the Ising spin to $\alpha = \bra{\uparrow} J^z \ket{\uparrow} = -\bra{\downarrow} J^z \ket{\downarrow} = \num{5.51}$.
For simplicity, the Ho magnetic moments are taken to be uniform. In previous work~\cite{dollbergEffectIntrinsicQuantum2022b}, we accounted for site-dependent variations, but the differences in moment size were small and contributed little compared to the dominant energetic effect. Given the larger energy scales in the present setting, such variations are expected to be even less relevant here.

The last term in \eqref{eq:MC-hamiltonian-LiHoErF4} plays only a secondary role in the present work. 
It is retained primarily to fix the absolute scale of the critical temperature, ensuring that the 
undiluted system reproduces the experimental value 
$T_c(x=1,B_x=0)=\qty{1.53}{\kelvin}$. 
Its contribution to the \emph{dependence} of the critical temperature on the Ho concentration $x$ 
is comparatively small, and previous work showed that it has only a minor effect on the slope 
of $T_c(x)$ relative to the mechanism discussed here~\cite{dollbergEffectIntrinsicQuantum2022b}. 
Accordingly, this term should be viewed as a quantitative baseline correction rather than as a key 
ingredient controlling the concentration dependence.

For completeness, we briefly summarize the origin and modeling of this interaction. 
The last term in \eqref{eq:MC-hamiltonian-LiHoErF4} describes emergent three-body interactions between 
Ho ions that arise when off-diagonal components of the dipolar interaction are incorporated 
perturbatively into the low-energy model. 
Starting from the full microscopic Hamiltonian, a Schrieffer--Wolff transformation is applied prior 
to projection, integrating out virtual excitations into higher crystal-field levels. 
To second order in the interaction, this procedure generates effective multispin terms acting 
within the low-energy Ising subspace. 
In particular, virtual processes in which two spins interact indirectly via a third spin that is 
excited and deexcited by off-diagonal dipolar couplings give rise to an effective interaction of 
the form $\propto \sum_k \epsilon^{\mathrm{Ho}}_k \left[ V_{ik}^{xz} V_{kj}^{xz} + V_{ik}^{yz} V_{kj}^{yz} \right]$. In light of the limited quantitative impact of this term on this work, we replace the site-dependent occupancy factor with its average value $\epsilon^{\mathrm{Ho}}_k\rightarrow \langle \epsilon^{\mathrm{Ho}}_k \rangle = x$, 
which corresponds to the Ho concentration. 
Furthermore, although the coefficient of this three-body term can be obtained microscopically from the 
Schrieffer--Wolff transformation, in the present work it is treated phenomenologically and fixed to $K=4.7\times10^{-5}a^6\,\mathrm{K}$, where $a=5.175\,\text{\AA}$ is the length of the \lhf{} unit cell, to reproduce the experimental critical temperature of the pure compound. 
The resulting value corresponds to a modest renormalization of the microscopic estimate, reflecting the remaining quantitative discrepancy between the perturbative derivation and experiment \cite{dollbergLiHoF4SpinhalfNonstandard2024}.

\begin{figure}[h!]
	\centering
	\includegraphics[width=\linewidth]{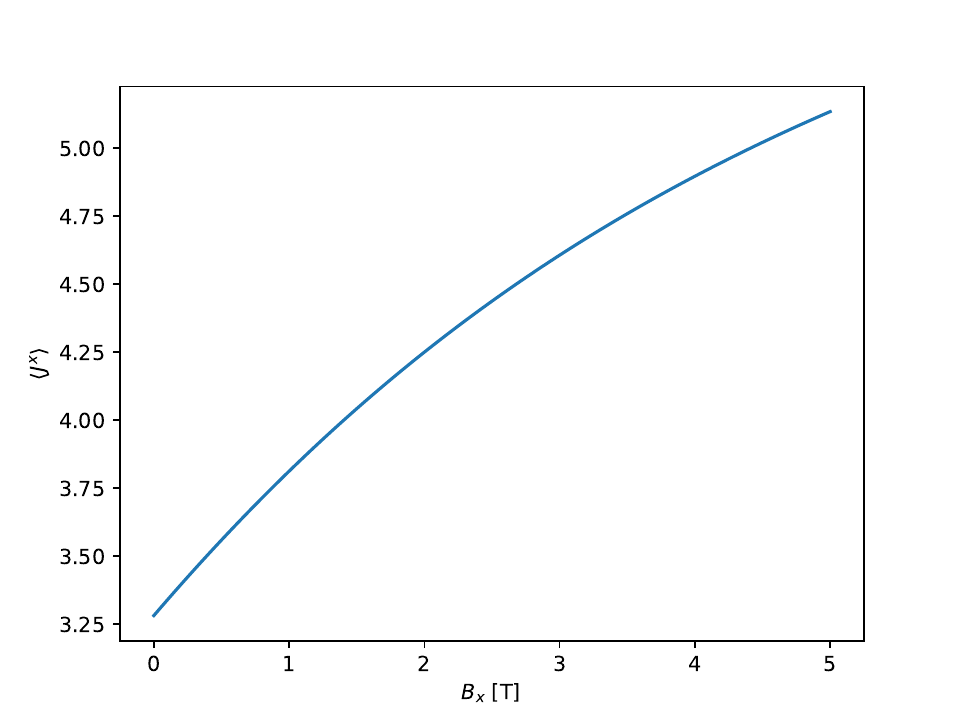}
	\caption{The expectation value $\langle J^x \rangle$ of an Er ion, calculated at $T=0$ under the crystal-field potential of \lef{}, as a function of a magnetic field applied along the $x$-axis. The field is always finite, approaching zero from above ($B_x \rightarrow 0^+$), to ensure a consistent symmetry-breaking direction. The same behavior is observed for a field applied along any arbitrary direction within the $x$–$y$ plane.}
	\label{fig:Er-Jx-vs-Bx}
\end{figure}

As for the Er ions, they are modeled as classical $XY$ spins with a fixed magnitude of $J_{xy}=\num{3.49}$. Notably, the Er spins exhibit considerable variations in their magnetic moments depending on the applied field, as illustrated in Fig.~\ref{fig:Er-Jx-vs-Bx}. The value we selected for the simulation corresponds to the typical nonzero internal transverse field found in Ref.~\cite{dollbergEffectIntrinsicQuantum2022b} (see Fig.~4 therein).

\section{Numerical Results}\label{app:numerical-res}
We present the simulation parameters and results in Table~\ref{table:sim-params-LiHoErF4}. Equilibration times are determined via logarithmic binning: observables are averaged over successively doubled time windows, and equilibrium is assumed once averages over three consecutive bins agree within error bars~\cite{katzgraberUniversalityThreedimensionalIsing2006,dollbergEffectIntrinsicQuantum2022b,dollbergLiHoF4SpinhalfNonstandard2024}.
\begin{table}[h!]
	\caption{Simulation parameters and results for Fig.~\ref{fig:LiHoErF4-x-T-phase-diagram}: $\mathrm{LiHo_{x}}R\mathrm{_{1-x}F_{4}}$ ($R$=Er,Y) for different dilutions $x\leq 1$ and system sizes $L$. The equilibration/measurement times are $2^b$ Monte Carlo sweeps. $T_\text{min}$ ($T_\text{max}$) is the lowest (highest) temperature used and $N_T$ is the number of temperatures. $N_{sa}$ is the number of independent runs. $T_c$ is the transition temperature, and $\nu$ is the critical exponent that describes the divergence of the correlation length. Temperatures are measured in Kelvin.}
	\label{table:sim-params-LiHoErF4}
	\centering
\begin{tabular}{ c c c c c c c c c c c c } 
    \hline\hline
    R & $x$ & L & b & $T_{\mathrm{min}}$ & $T_{\mathrm{max}}$ & $N_T$ & $N_{sa}$ & $T_c$ & $\nu$ \\ [0.5ex] 
    \hline
        & \multirow{2}{*}{1.0} & 7,8,9 & 12 & 1.326 & 1.726 & 24 & 50 & \multirow{2}{*}{1.5206(5)} & \multirow{2}{*}{0.630(9)} \\
        & & 10,11 & 14 & 1.319 & 1.719 & 24 & 50 & & \\
    \hline 
    \multirow{17}{*}{Er} & \multirow{2}{*}{0.86} & 7,8,9 & 13 & 1.048 & 1.448 & 24 & 175 & \multirow{2}{*}{1.2613(9)} & \multirow{2}{*}{0.65(2)} \\ 
        & & 10,11 & 14 & 1.066 & 1.466 & 24 & 175 &  &  \\ \cline{2-10}
        & \multirow{4}{*}{0.79} & 7,8 & 12 & 0.929 & 1.329 & 24 & 235 & \multirow{4}{*}{1.122(1)} & \multirow{4}{*}{0.68(2)} \\
        & & 9 & 14 & 0.929 & 1.329 & 24 & 235 &  &  \\
        & & 10 & 14 & 0.91 & 1.31 & 24 & 235 &  &  \\
        & & 11 & 15 & 0.91 & 1.31 & 24 & 235 &  &  \\ \cline{2-10}
        & \multirow{4}{*}{0.765} & 7,8 & 12 & 0.842 & 1.242 & 24 & 260 & \multirow{4}{*}{1.072(2)} & \multirow{4}{*}{0.68(3)} \\
        & & 9 & 14 & 0.842 & 1.242 & 24 & 260 &  &  \\
        & & 10 & 14 & 0.855 & 1.255 & 24 & 260 &  &  \\
        & & 11 & 15 & 0.855 & 1.255 & 24 & 260 &  &  \\ \cline{2-10}
        & \multirow{3}{*}{0.7} & 7,8 & 13 & 0.705 & 1.105 & 24 & 480 & \multirow{3}{*}{0.925(1)} & \multirow{3}{*}{0.71(3)} \\
        & & 9 & 14 & 0.705 & 1.105 & 24 & 480 &  &  \\
        & & 10,11 & 14 & 0.711 & 1.111 & 24 & 480 &  &  \\ \cline{2-10}
        & \multirow{4}{*}{0.675} & 7,8 & 13 & 0.636 & 1.036 & 24 & 520 & \multirow{4}{*}{0.868(1)} & \multirow{4}{*}{0.71(2)} \\
        & & 9 & 14 & 0.636 & 1.036 & 24 & 520 &  &  \\
        & & 10 & 15 & 0.663 & 1.063 & 24 & 520 & & \\
        & & 11 & 15 & 0.663 & 1.063 & 24 & 570 & & \\
    \hline
    \multirow{6}{*}{Y} & \multirow{3}{*}{0.83} & 7,8 & 13 & 1.073 & 1.473 & 24 & 200 & \multirow{3}{*}{1.2785(9)} & \multirow{3}{*}{0.64(2)} \\ 
        & & 9 & 15 & 1.073 & 1.473 & 24 & 200 & & \\
        & & 10,11 & 14 & 1.078 & 1.478 & 24 & 200 & & \\ \cline{2-10} 
        & \multirow{3}{*}{0.67} & 7,8,9 & 13 & 0.8 & 1.2 & 24 & 520 & \multirow{3}{*}{1.0262(8)} & \multirow{3}{*}{0.68(2)} \\
        & & 10 & 14 & 0.819 & 1.219 & 24 & 520 & & \\
        & & 11 & 15 & 0.819 & 1.219 & 24 & 520 & & \\
    \hline\hline
\end{tabular}

\end{table}


\begin{thebibliography}{57}%
\makeatletter
\providecommand \@ifxundefined [1]{%
 \@ifx{#1\undefined}
}%
\providecommand \@ifnum [1]{%
 \ifnum #1\expandafter \@firstoftwo
 \else \expandafter \@secondoftwo
 \fi
}%
\providecommand \@ifx [1]{%
 \ifx #1\expandafter \@firstoftwo
 \else \expandafter \@secondoftwo
 \fi
}%
\providecommand \natexlab [1]{#1}%
\providecommand \enquote  [1]{``#1''}%
\providecommand \bibnamefont  [1]{#1}%
\providecommand \bibfnamefont [1]{#1}%
\providecommand \citenamefont [1]{#1}%
\providecommand \href@noop [0]{\@secondoftwo}%
\providecommand \href [0]{\begingroup \@sanitize@url \@href}%
\providecommand \@href[1]{\@@startlink{#1}\@@href}%
\providecommand \@@href[1]{\endgroup#1\@@endlink}%
\providecommand \@sanitize@url [0]{\catcode `\\12\catcode `\$12\catcode
  `\&12\catcode `\#12\catcode `\^12\catcode `\_12\catcode `\%12\relax}%
\providecommand \@@startlink[1]{}%
\providecommand \@@endlink[0]{}%
\providecommand \url  [0]{\begingroup\@sanitize@url \@url }%
\providecommand \@url [1]{\endgroup\@href {#1}{\urlprefix }}%
\providecommand \urlprefix  [0]{URL }%
\providecommand \Eprint [0]{\href }%
\providecommand \doibase [0]{https://doi.org/}%
\providecommand \selectlanguage [0]{\@gobble}%
\providecommand \bibinfo  [0]{\@secondoftwo}%
\providecommand \bibfield  [0]{\@secondoftwo}%
\providecommand \translation [1]{[#1]}%
\providecommand \BibitemOpen [0]{}%
\providecommand \bibitemStop [0]{}%
\providecommand \bibitemNoStop [0]{.\EOS\space}%
\providecommand \EOS [0]{\spacefactor3000\relax}%
\providecommand \BibitemShut  [1]{\csname bibitem#1\endcsname}%
\let\auto@bib@innerbib\@empty
\bibitem [{\citenamefont {Alben}\ \emph {et~al.}(1978)\citenamefont {Alben},
  \citenamefont {Becker},\ and\ \citenamefont
  {Chi}}]{albenRandomAnisotropyAmorphous1978}%
  \BibitemOpen
  \bibfield  {author} {\bibinfo {author} {\bibfnamefont {R.}~\bibnamefont
  {Alben}}, \bibinfo {author} {\bibfnamefont {J.~J.}\ \bibnamefont {Becker}},\
  and\ \bibinfo {author} {\bibfnamefont {M.~C.}\ \bibnamefont {Chi}},\
  }\bibfield  {title} {\bibinfo {title} {Random anisotropy in amorphous
  ferromagnets},\ }\href {https://doi.org/10.1063/1.324881} {\bibfield
  {journal} {\bibinfo  {journal} {Journal of Applied Physics}\ }\textbf
  {\bibinfo {volume} {49}},\ \bibinfo {pages} {1653} (\bibinfo {year}
  {1978})}\BibitemShut {NoStop}%
\bibitem [{\citenamefont {Fishman}\ and\ \citenamefont
  {Aharony}(1978)}]{fishmanPhaseDiagramsMulticritical1978}%
  \BibitemOpen
  \bibfield  {author} {\bibinfo {author} {\bibfnamefont {S.}~\bibnamefont
  {Fishman}}\ and\ \bibinfo {author} {\bibfnamefont {A.}~\bibnamefont
  {Aharony}},\ }\bibfield  {title} {\bibinfo {title} {Phase diagrams and
  multicritical points in randomly mixed magnets. {{I}}. {{Mixed}}
  anisotropies},\ }\href {https://doi.org/10.1103/PhysRevB.18.3507} {\bibfield
  {journal} {\bibinfo  {journal} {Physical Review B}\ }\textbf {\bibinfo
  {volume} {18}},\ \bibinfo {pages} {3507} (\bibinfo {year}
  {1978})}\BibitemShut {NoStop}%
\bibitem [{\citenamefont {Wong}\ \emph {et~al.}(1980)\citenamefont {Wong},
  \citenamefont {Horn}, \citenamefont {Birgeneau}, \citenamefont {Safinya},\
  and\ \citenamefont {Shirane}}]{wongCompetingOrderParameters1980}%
  \BibitemOpen
  \bibfield  {author} {\bibinfo {author} {\bibfnamefont {P.-z.}\ \bibnamefont
  {Wong}}, \bibinfo {author} {\bibfnamefont {P.~M.}\ \bibnamefont {Horn}},
  \bibinfo {author} {\bibfnamefont {R.~J.}\ \bibnamefont {Birgeneau}}, \bibinfo
  {author} {\bibfnamefont {C.~R.}\ \bibnamefont {Safinya}},\ and\ \bibinfo
  {author} {\bibfnamefont {G.}~\bibnamefont {Shirane}},\ }\bibfield  {title}
  {\bibinfo {title} {Competing {{Order Parameters}} in {{Quenched Random
  Alloys}}:
  {{Fe}}\textsubscript{1-x}{{Co}}\textsubscript{x}{{Cl}}\textsubscript{2}},\
  }\href {https://doi.org/10.1103/PhysRevLett.45.1974} {\bibfield  {journal}
  {\bibinfo  {journal} {Physical Review Letters}\ }\textbf {\bibinfo {volume}
  {45}},\ \bibinfo {pages} {1974} (\bibinfo {year} {1980})}\BibitemShut
  {NoStop}%
\bibitem [{\citenamefont
  {Mukamel}(1981)}]{mukamelPhaseDiagramsMulticritical1981}%
  \BibitemOpen
  \bibfield  {author} {\bibinfo {author} {\bibfnamefont {D.}~\bibnamefont
  {Mukamel}},\ }\bibfield  {title} {\bibinfo {title} {Phase {{Diagrams}} and
  {{Multicritical Points}} in {{Randomly Mixed Alloys}}},\ }\href
  {https://doi.org/10.1103/PhysRevLett.46.845} {\bibfield  {journal} {\bibinfo
  {journal} {Physical Review Letters}\ }\textbf {\bibinfo {volume} {46}},\
  \bibinfo {pages} {845} (\bibinfo {year} {1981})}\BibitemShut {NoStop}%
\bibitem [{\citenamefont {Wong}\ \emph {et~al.}(1983)\citenamefont {Wong},
  \citenamefont {Horn}, \citenamefont {Birgeneau},\ and\ \citenamefont
  {Shirane}}]{wongFe1xCoxCl2CompetingAnisotropies1983}%
  \BibitemOpen
  \bibfield  {author} {\bibinfo {author} {\bibfnamefont {P.}~\bibnamefont
  {Wong}}, \bibinfo {author} {\bibfnamefont {P.~M.}\ \bibnamefont {Horn}},
  \bibinfo {author} {\bibfnamefont {R.~J.}\ \bibnamefont {Birgeneau}},\ and\
  \bibinfo {author} {\bibfnamefont {G.}~\bibnamefont {Shirane}},\ }\bibfield
  {title} {\bibinfo {title}
  {{Fe}\textsubscript{1-x}{Co}\textsubscript{x}{Cl}\textsubscript{2}:
  {{Competing}} anisotropies and random molecular fields},\ }\href
  {https://doi.org/10.1103/PhysRevB.27.428} {\bibfield  {journal} {\bibinfo
  {journal} {Physical Review B}\ }\textbf {\bibinfo {volume} {27}},\ \bibinfo
  {pages} {428} (\bibinfo {year} {1983})}\BibitemShut {NoStop}%
\bibitem [{\citenamefont {DeFotis}\ \emph {et~al.}(1984)\citenamefont
  {DeFotis}, \citenamefont {Pohl}, \citenamefont {Pugh},\ and\ \citenamefont
  {Sinn}}]{defotisMagneticPhaseDiagram1984a}%
  \BibitemOpen
  \bibfield  {author} {\bibinfo {author} {\bibfnamefont {G.~C.}\ \bibnamefont
  {DeFotis}}, \bibinfo {author} {\bibfnamefont {C.}~\bibnamefont {Pohl}},
  \bibinfo {author} {\bibfnamefont {S.~A.}\ \bibnamefont {Pugh}},\ and\
  \bibinfo {author} {\bibfnamefont {E.}~\bibnamefont {Sinn}},\ }\bibfield
  {title} {\bibinfo {title} {Magnetic phase diagram and spin glass behavior of
  {{Fe}}\textsubscript{1-x}{{Mn}}\textsubscript{x}{{Cl}}\textsubscript{2}{$\cdot$}{{2H\textsubscript{2}O}}},\
  }\href {https://doi.org/10.1063/1.446973} {\bibfield  {journal} {\bibinfo
  {journal} {The Journal of Chemical Physics}\ }\textbf {\bibinfo {volume}
  {80}},\ \bibinfo {pages} {2079} (\bibinfo {year} {1984})}\BibitemShut
  {NoStop}%
\bibitem [{\citenamefont {Ibarra}\ \emph {et~al.}(1991)\citenamefont {Ibarra},
  \citenamefont {Morellon}, \citenamefont {Algarabel},\ and\ \citenamefont
  {Moze}}]{ibarraSingleionCompetingMagnetic1991}%
  \BibitemOpen
  \bibfield  {author} {\bibinfo {author} {\bibfnamefont {M.~R.}\ \bibnamefont
  {Ibarra}}, \bibinfo {author} {\bibfnamefont {L.}~\bibnamefont {Morellon}},
  \bibinfo {author} {\bibfnamefont {P.~A.}\ \bibnamefont {Algarabel}},\ and\
  \bibinfo {author} {\bibfnamefont {O.}~\bibnamefont {Moze}},\ }\bibfield
  {title} {\bibinfo {title} {Single-ion competing magnetic anisotropies in
  {${\mathrm{Pr}}_{\mathit{x}}$}{${\mathrm{Nd}}_{1\mathrm{\ensuremath{-}}\mathit{x}}$}{${\mathrm{Co}}_{5}$}
  intermetallic compounds},\ }\href {https://doi.org/10.1103/PhysRevB.44.9368}
  {\bibfield  {journal} {\bibinfo  {journal} {Physical Review B}\ }\textbf
  {\bibinfo {volume} {44}},\ \bibinfo {pages} {9368} (\bibinfo {year}
  {1991})}\BibitemShut {NoStop}%
\bibitem [{\citenamefont {Katsumata}\ \emph {et~al.}(1992)\citenamefont
  {Katsumata}, \citenamefont {Shapiro}, \citenamefont {Matsuda}, \citenamefont
  {Shirane},\ and\ \citenamefont
  {Tuchendler}}]{katsumataSimultaneousOrderingOrthogonal1992}%
  \BibitemOpen
  \bibfield  {author} {\bibinfo {author} {\bibfnamefont {K.}~\bibnamefont
  {Katsumata}}, \bibinfo {author} {\bibfnamefont {S.~M.}\ \bibnamefont
  {Shapiro}}, \bibinfo {author} {\bibfnamefont {M.}~\bibnamefont {Matsuda}},
  \bibinfo {author} {\bibfnamefont {G.}~\bibnamefont {Shirane}},\ and\ \bibinfo
  {author} {\bibfnamefont {J.}~\bibnamefont {Tuchendler}},\ }\bibfield  {title}
  {\bibinfo {title} {Simultaneous ordering of orthogonal spin components in a
  random magnet with competing anisotropies},\ }\href
  {https://doi.org/10.1103/PhysRevB.46.14906} {\bibfield  {journal} {\bibinfo
  {journal} {Physical Review B}\ }\textbf {\bibinfo {volume} {46}},\ \bibinfo
  {pages} {14906} (\bibinfo {year} {1992})}\BibitemShut {NoStop}%
\bibitem [{\citenamefont {Pirogov}\ \emph {et~al.}(2009)\citenamefont
  {Pirogov}, \citenamefont {Park}, \citenamefont {Ermolenko}, \citenamefont
  {Korolev}, \citenamefont {Kuchin}, \citenamefont {Lee}, \citenamefont {Choi},
  \citenamefont {Park}, \citenamefont {Ranot}, \citenamefont {Yi},
  \citenamefont {Gerasimov}, \citenamefont {Dorofeev}, \citenamefont
  {Vokhmyanin}, \citenamefont {Podlesnyak},\ and\ \citenamefont
  {Swainson}}]{pirogovTbxEr1xNi5CompoundsIdeal2009}%
  \BibitemOpen
  \bibfield  {author} {\bibinfo {author} {\bibfnamefont {A.~N.}\ \bibnamefont
  {Pirogov}}, \bibinfo {author} {\bibfnamefont {J.-G.}\ \bibnamefont {Park}},
  \bibinfo {author} {\bibfnamefont {A.~S.}\ \bibnamefont {Ermolenko}}, \bibinfo
  {author} {\bibfnamefont {A.~V.}\ \bibnamefont {Korolev}}, \bibinfo {author}
  {\bibfnamefont {A.~G.}\ \bibnamefont {Kuchin}}, \bibinfo {author}
  {\bibfnamefont {S.}~\bibnamefont {Lee}}, \bibinfo {author} {\bibfnamefont
  {Y.~N.}\ \bibnamefont {Choi}}, \bibinfo {author} {\bibfnamefont
  {J.}~\bibnamefont {Park}}, \bibinfo {author} {\bibfnamefont {M.}~\bibnamefont
  {Ranot}}, \bibinfo {author} {\bibfnamefont {J.}~\bibnamefont {Yi}}, \bibinfo
  {author} {\bibfnamefont {E.~G.}\ \bibnamefont {Gerasimov}}, \bibinfo {author}
  {\bibfnamefont {{\relax Yu}.~A.}\ \bibnamefont {Dorofeev}}, \bibinfo {author}
  {\bibfnamefont {A.~P.}\ \bibnamefont {Vokhmyanin}}, \bibinfo {author}
  {\bibfnamefont {A.~A.}\ \bibnamefont {Podlesnyak}},\ and\ \bibinfo {author}
  {\bibfnamefont {I.~P.}\ \bibnamefont {Swainson}},\ }\bibfield  {title}
  {\bibinfo {title}
  {{${\text{Tb}}_{x}{\text{Er}}_{1\ensuremath{-}x}{\text{Ni}}_{5}$} compounds:
  {{An}} ideal model system for competing {{Ising-}}{$XY$} anisotropy
  energies},\ }\href {https://doi.org/10.1103/PhysRevB.79.174412} {\bibfield
  {journal} {\bibinfo  {journal} {Physical Review B}\ }\textbf {\bibinfo
  {volume} {79}},\ \bibinfo {pages} {174412} (\bibinfo {year}
  {2009})}\BibitemShut {NoStop}%
\bibitem [{\citenamefont {Perez}\ \emph {et~al.}(2015)\citenamefont {Perez},
  \citenamefont {Borisov}, \citenamefont {Johnson}, \citenamefont {Stanescu},
  \citenamefont {Trappen}, \citenamefont {Holcomb}, \citenamefont {Lederman},
  \citenamefont {Fitzsimmons}, \citenamefont {Aczel},\ and\ \citenamefont
  {Hong}}]{perezPhaseDiagramThreeDimensional2015}%
  \BibitemOpen
  \bibfield  {author} {\bibinfo {author} {\bibfnamefont {F.~A.}\ \bibnamefont
  {Perez}}, \bibinfo {author} {\bibfnamefont {P.}~\bibnamefont {Borisov}},
  \bibinfo {author} {\bibfnamefont {T.~A.}\ \bibnamefont {Johnson}}, \bibinfo
  {author} {\bibfnamefont {T.~D.}\ \bibnamefont {Stanescu}}, \bibinfo {author}
  {\bibfnamefont {R.}~\bibnamefont {Trappen}}, \bibinfo {author} {\bibfnamefont
  {M.~B.}\ \bibnamefont {Holcomb}}, \bibinfo {author} {\bibfnamefont
  {D.}~\bibnamefont {Lederman}}, \bibinfo {author} {\bibfnamefont {M.~R.}\
  \bibnamefont {Fitzsimmons}}, \bibinfo {author} {\bibfnamefont {A.~A.}\
  \bibnamefont {Aczel}},\ and\ \bibinfo {author} {\bibfnamefont
  {T.}~\bibnamefont {Hong}},\ }\bibfield  {title} {\bibinfo {title} {Phase
  {{Diagram}} of a {{Three-Dimensional Antiferromagnet}} with {{Random Magnetic
  Anisotropy}}},\ }\href {https://doi.org/10.1103/PhysRevLett.114.097201}
  {\bibfield  {journal} {\bibinfo  {journal} {Physical Review Letters}\
  }\textbf {\bibinfo {volume} {114}},\ \bibinfo {pages} {097201} (\bibinfo
  {year} {2015})}\BibitemShut {NoStop}%
\bibitem [{\citenamefont {Bhutani}\ \emph {et~al.}(2020)\citenamefont
  {Bhutani}, \citenamefont {Zuo}, \citenamefont {McAuliffe}, \citenamefont
  {{dela Cruz}},\ and\ \citenamefont
  {Shoemaker}}]{bhutaniStrongAnisotropyMixed2020a}%
  \BibitemOpen
  \bibfield  {author} {\bibinfo {author} {\bibfnamefont {A.}~\bibnamefont
  {Bhutani}}, \bibinfo {author} {\bibfnamefont {J.~L.}\ \bibnamefont {Zuo}},
  \bibinfo {author} {\bibfnamefont {R.~D.}\ \bibnamefont {McAuliffe}}, \bibinfo
  {author} {\bibfnamefont {C.~R.}\ \bibnamefont {{dela Cruz}}},\ and\ \bibinfo
  {author} {\bibfnamefont {D.~P.}\ \bibnamefont {Shoemaker}},\ }\bibfield
  {title} {\bibinfo {title} {Strong anisotropy in the mixed antiferromagnetic
  system
  {${\mathrm{Mn}}_{1\ensuremath{-}x}{\mathrm{Fe}}_{x}{\mathrm{PSe}}_{3}$}},\
  }\href {https://doi.org/10.1103/PhysRevMaterials.4.034411} {\bibfield
  {journal} {\bibinfo  {journal} {Physical Review Materials}\ }\textbf
  {\bibinfo {volume} {4}},\ \bibinfo {pages} {034411} (\bibinfo {year}
  {2020})}\BibitemShut {NoStop}%
\bibitem [{\citenamefont {Fogh}\ \emph {et~al.}(2023)\citenamefont {Fogh},
  \citenamefont {Klemke}, \citenamefont {Reehuis}, \citenamefont {Bourges},
  \citenamefont {Niedermayer}, \citenamefont {{Holm-Dahlin}}, \citenamefont
  {Zaharko}, \citenamefont {Schefer}, \citenamefont {Kristensen}, \citenamefont
  {S{\o}rensen}, \citenamefont {Paeckel}, \citenamefont {Pedersen},
  \citenamefont {Hansen}, \citenamefont {Pages}, \citenamefont {Moerner},
  \citenamefont {Meucci}, \citenamefont {Soh}, \citenamefont {Bombardi},
  \citenamefont {Vaknin}, \citenamefont {R{\o}nnow}, \citenamefont
  {Sylju{\aa}sen}, \citenamefont {Christensen},\ and\ \citenamefont
  {{Toft-Petersen}}}]{foghTuningMagnetoelectricityMixedanisotropy2023}%
  \BibitemOpen
  \bibfield  {author} {\bibinfo {author} {\bibfnamefont {E.}~\bibnamefont
  {Fogh}}, \bibinfo {author} {\bibfnamefont {B.}~\bibnamefont {Klemke}},
  \bibinfo {author} {\bibfnamefont {M.}~\bibnamefont {Reehuis}}, \bibinfo
  {author} {\bibfnamefont {P.}~\bibnamefont {Bourges}}, \bibinfo {author}
  {\bibfnamefont {C.}~\bibnamefont {Niedermayer}}, \bibinfo {author}
  {\bibfnamefont {S.}~\bibnamefont {{Holm-Dahlin}}}, \bibinfo {author}
  {\bibfnamefont {O.}~\bibnamefont {Zaharko}}, \bibinfo {author} {\bibfnamefont
  {J.}~\bibnamefont {Schefer}}, \bibinfo {author} {\bibfnamefont {A.~B.}\
  \bibnamefont {Kristensen}}, \bibinfo {author} {\bibfnamefont {M.~K.}\
  \bibnamefont {S{\o}rensen}}, \bibinfo {author} {\bibfnamefont
  {S.}~\bibnamefont {Paeckel}}, \bibinfo {author} {\bibfnamefont {K.~S.}\
  \bibnamefont {Pedersen}}, \bibinfo {author} {\bibfnamefont {R.~E.}\
  \bibnamefont {Hansen}}, \bibinfo {author} {\bibfnamefont {A.}~\bibnamefont
  {Pages}}, \bibinfo {author} {\bibfnamefont {K.~K.}\ \bibnamefont {Moerner}},
  \bibinfo {author} {\bibfnamefont {G.}~\bibnamefont {Meucci}}, \bibinfo
  {author} {\bibfnamefont {J.-R.}\ \bibnamefont {Soh}}, \bibinfo {author}
  {\bibfnamefont {A.}~\bibnamefont {Bombardi}}, \bibinfo {author}
  {\bibfnamefont {D.}~\bibnamefont {Vaknin}}, \bibinfo {author} {\bibfnamefont
  {{\relax Henrik}.~M.}\ \bibnamefont {R{\o}nnow}}, \bibinfo {author}
  {\bibfnamefont {O.~F.}\ \bibnamefont {Sylju{\aa}sen}}, \bibinfo {author}
  {\bibfnamefont {N.~B.}\ \bibnamefont {Christensen}},\ and\ \bibinfo {author}
  {\bibfnamefont {R.}~\bibnamefont {{Toft-Petersen}}},\ }\bibfield  {title}
  {\bibinfo {title} {Tuning magnetoelectricity in a mixed-anisotropy
  antiferromagnet},\ }\href {https://doi.org/10.1038/s41467-023-39128-7}
  {\bibfield  {journal} {\bibinfo  {journal} {Nature Communications}\ }\textbf
  {\bibinfo {volume} {14}},\ \bibinfo {pages} {3408} (\bibinfo {year}
  {2023})}\BibitemShut {NoStop}%
\bibitem [{\citenamefont {Wu}\ \emph {et~al.}(1991)\citenamefont {Wu},
  \citenamefont {Ellman}, \citenamefont {Rosenbaum}, \citenamefont {Aeppli},\
  and\ \citenamefont {Reich}}]{wuClassicalQuantumGlass1991a}%
  \BibitemOpen
  \bibfield  {author} {\bibinfo {author} {\bibfnamefont {W.}~\bibnamefont
  {Wu}}, \bibinfo {author} {\bibfnamefont {B.}~\bibnamefont {Ellman}}, \bibinfo
  {author} {\bibfnamefont {T.~F.}\ \bibnamefont {Rosenbaum}}, \bibinfo {author}
  {\bibfnamefont {G.}~\bibnamefont {Aeppli}},\ and\ \bibinfo {author}
  {\bibfnamefont {D.~H.}\ \bibnamefont {Reich}},\ }\bibfield  {title} {\bibinfo
  {title} {From classical to quantum glass},\ }\href
  {https://doi.org/10.1103/PhysRevLett.67.2076} {\bibfield  {journal} {\bibinfo
   {journal} {Physical Review Letters}\ }\textbf {\bibinfo {volume} {67}},\
  \bibinfo {pages} {2076} (\bibinfo {year} {1991})}\BibitemShut {NoStop}%
\bibitem [{\citenamefont {Brooke}\ \emph {et~al.}(1999)\citenamefont {Brooke},
  \citenamefont {Bitko}, \citenamefont {Rosenbaum},\ and\ \citenamefont
  {Aeppli}}]{brookeQuantumAnnealingDisordered1999}%
  \BibitemOpen
  \bibfield  {author} {\bibinfo {author} {\bibfnamefont {J.}~\bibnamefont
  {Brooke}}, \bibinfo {author} {\bibfnamefont {D.}~\bibnamefont {Bitko}},
  \bibinfo {author} {\bibfnamefont {T.~F.}\ \bibnamefont {Rosenbaum}},\ and\
  \bibinfo {author} {\bibfnamefont {G.}~\bibnamefont {Aeppli}},\ }\bibfield
  {title} {\bibinfo {title} {Quantum {{Annealing}} of a {{Disordered
  Magnet}}},\ }\href {https://doi.org/10.1126/science.284.5415.779} {\bibfield
  {journal} {\bibinfo  {journal} {Science}\ }\textbf {\bibinfo {volume}
  {284}},\ \bibinfo {pages} {779} (\bibinfo {year} {1999})}\BibitemShut
  {NoStop}%
\bibitem [{\citenamefont {Schechter}\ and\ \citenamefont
  {Laflorencie}(2006)}]{schechterQuantumSpinGlass2006}%
  \BibitemOpen
  \bibfield  {author} {\bibinfo {author} {\bibfnamefont {M.}~\bibnamefont
  {Schechter}}\ and\ \bibinfo {author} {\bibfnamefont {N.}~\bibnamefont
  {Laflorencie}},\ }\bibfield  {title} {\bibinfo {title} {Quantum {{Spin
  Glass}} and the {{Dipolar Interaction}}},\ }\href
  {https://doi.org/10.1103/PhysRevLett.97.137204} {\bibfield  {journal}
  {\bibinfo  {journal} {Physical Review Letters}\ }\textbf {\bibinfo {volume}
  {97}},\ \bibinfo {pages} {137204} (\bibinfo {year} {2006})}\BibitemShut
  {NoStop}%
\bibitem [{\citenamefont {Kraemer}\ \emph {et~al.}(2012)\citenamefont
  {Kraemer}, \citenamefont {Nikseresht}, \citenamefont {Piatek}, \citenamefont
  {Tsyrulin}, \citenamefont {Piazza}, \citenamefont {Kiefer}, \citenamefont
  {Klemke}, \citenamefont {Rosenbaum}, \citenamefont {Aeppli}, \citenamefont
  {Gannarelli}, \citenamefont {Prokes}, \citenamefont {Podlesnyak},
  \citenamefont {Str{\"a}ssle}, \citenamefont {Keller}, \citenamefont
  {Zaharko}, \citenamefont {Kr{\"a}mer},\ and\ \citenamefont
  {R{\o}nnow}}]{kraemerDipolarAntiferromagnetismQuantum2012}%
  \BibitemOpen
  \bibfield  {author} {\bibinfo {author} {\bibfnamefont {C.}~\bibnamefont
  {Kraemer}}, \bibinfo {author} {\bibfnamefont {N.}~\bibnamefont {Nikseresht}},
  \bibinfo {author} {\bibfnamefont {J.~O.}\ \bibnamefont {Piatek}}, \bibinfo
  {author} {\bibfnamefont {N.}~\bibnamefont {Tsyrulin}}, \bibinfo {author}
  {\bibfnamefont {B.~D.}\ \bibnamefont {Piazza}}, \bibinfo {author}
  {\bibfnamefont {K.}~\bibnamefont {Kiefer}}, \bibinfo {author} {\bibfnamefont
  {B.}~\bibnamefont {Klemke}}, \bibinfo {author} {\bibfnamefont {T.~F.}\
  \bibnamefont {Rosenbaum}}, \bibinfo {author} {\bibfnamefont {G.}~\bibnamefont
  {Aeppli}}, \bibinfo {author} {\bibfnamefont {C.}~\bibnamefont {Gannarelli}},
  \bibinfo {author} {\bibfnamefont {K.}~\bibnamefont {Prokes}}, \bibinfo
  {author} {\bibfnamefont {A.}~\bibnamefont {Podlesnyak}}, \bibinfo {author}
  {\bibfnamefont {T.}~\bibnamefont {Str{\"a}ssle}}, \bibinfo {author}
  {\bibfnamefont {L.}~\bibnamefont {Keller}}, \bibinfo {author} {\bibfnamefont
  {O.}~\bibnamefont {Zaharko}}, \bibinfo {author} {\bibfnamefont {K.~W.}\
  \bibnamefont {Kr{\"a}mer}},\ and\ \bibinfo {author} {\bibfnamefont {H.~M.}\
  \bibnamefont {R{\o}nnow}},\ }\bibfield  {title} {\bibinfo {title} {Dipolar
  {{Antiferromagnetism}} and {{Quantum Criticality}} in
  {{LiErF\textsubscript{4}}}},\ }\href
  {https://doi.org/10.1126/science.1221878} {\bibfield  {journal} {\bibinfo
  {journal} {Science}\ }\textbf {\bibinfo {volume} {336}},\ \bibinfo {pages}
  {1416} (\bibinfo {year} {2012})}\BibitemShut {NoStop}%
\bibitem [{\citenamefont {Silevitch}\ \emph {et~al.}(2007)\citenamefont
  {Silevitch}, \citenamefont {Bitko}, \citenamefont {Brooke}, \citenamefont
  {Ghosh}, \citenamefont {Aeppli},\ and\ \citenamefont
  {Rosenbaum}}]{silevitchFerromagnetContinuouslyTunable2007}%
  \BibitemOpen
  \bibfield  {author} {\bibinfo {author} {\bibfnamefont {D.~M.}\ \bibnamefont
  {Silevitch}}, \bibinfo {author} {\bibfnamefont {D.}~\bibnamefont {Bitko}},
  \bibinfo {author} {\bibfnamefont {J.}~\bibnamefont {Brooke}}, \bibinfo
  {author} {\bibfnamefont {S.}~\bibnamefont {Ghosh}}, \bibinfo {author}
  {\bibfnamefont {G.}~\bibnamefont {Aeppli}},\ and\ \bibinfo {author}
  {\bibfnamefont {T.~F.}\ \bibnamefont {Rosenbaum}},\ }\bibfield  {title}
  {\bibinfo {title} {A ferromagnet in a continuously tunable random field},\
  }\href {https://doi.org/10.1038/nature06050} {\bibfield  {journal} {\bibinfo
  {journal} {Nature}\ }\textbf {\bibinfo {volume} {448}},\ \bibinfo {pages}
  {567} (\bibinfo {year} {2007})}\BibitemShut {NoStop}%
\bibitem [{\citenamefont {Liu}\ \emph {et~al.}(2023)\citenamefont {Liu},
  \citenamefont {Yuan}, \citenamefont {Dong}, \citenamefont {Lin},
  \citenamefont {V{\'i}llora}, \citenamefont {Qi}, \citenamefont {Zhao},
  \citenamefont {Shimamura}, \citenamefont {Ma}, \citenamefont {Wang},
  \citenamefont {Zhang},\ and\ \citenamefont
  {Li}}]{liuUltralowfieldMagnetocaloricMaterials2023}%
  \BibitemOpen
  \bibfield  {author} {\bibinfo {author} {\bibfnamefont {P.}~\bibnamefont
  {Liu}}, \bibinfo {author} {\bibfnamefont {D.}~\bibnamefont {Yuan}}, \bibinfo
  {author} {\bibfnamefont {C.}~\bibnamefont {Dong}}, \bibinfo {author}
  {\bibfnamefont {G.}~\bibnamefont {Lin}}, \bibinfo {author} {\bibfnamefont
  {E.~G.}\ \bibnamefont {V{\'i}llora}}, \bibinfo {author} {\bibfnamefont
  {J.}~\bibnamefont {Qi}}, \bibinfo {author} {\bibfnamefont {X.}~\bibnamefont
  {Zhao}}, \bibinfo {author} {\bibfnamefont {K.}~\bibnamefont {Shimamura}},
  \bibinfo {author} {\bibfnamefont {J.}~\bibnamefont {Ma}}, \bibinfo {author}
  {\bibfnamefont {J.}~\bibnamefont {Wang}}, \bibinfo {author} {\bibfnamefont
  {Z.}~\bibnamefont {Zhang}},\ and\ \bibinfo {author} {\bibfnamefont
  {B.}~\bibnamefont {Li}},\ }\bibfield  {title} {\bibinfo {title}
  {Ultralow-field magnetocaloric materials for compact magnetic
  refrigeration},\ }\href {https://doi.org/10.1038/s41427-023-00488-7}
  {\bibfield  {journal} {\bibinfo  {journal} {NPG Asia Materials}\ }\textbf
  {\bibinfo {volume} {15}},\ \bibinfo {pages} {41} (\bibinfo {year}
  {2023})}\BibitemShut {NoStop}%
\bibitem [{\citenamefont {Reich}\ \emph {et~al.}(1987)\citenamefont {Reich},
  \citenamefont {Rosenbaum},\ and\ \citenamefont
  {Aeppli}}]{reichGlassyRelaxationFreezing1987}%
  \BibitemOpen
  \bibfield  {author} {\bibinfo {author} {\bibfnamefont {D.~H.}\ \bibnamefont
  {Reich}}, \bibinfo {author} {\bibfnamefont {T.~F.}\ \bibnamefont
  {Rosenbaum}},\ and\ \bibinfo {author} {\bibfnamefont {G.}~\bibnamefont
  {Aeppli}},\ }\bibfield  {title} {\bibinfo {title} {Glassy relaxation without
  freezing in a random dipolar-coupled {{Ising}} magnet},\ }\href
  {https://doi.org/10.1103/physrevlett.59.1969} {\bibfield  {journal} {\bibinfo
   {journal} {Physical Review Letters}\ }\textbf {\bibinfo {volume} {59}},\
  \bibinfo {pages} {1969} (\bibinfo {year} {1987})}\BibitemShut {NoStop}%
\bibitem [{\citenamefont {Giraud}\ \emph {et~al.}(2001)\citenamefont {Giraud},
  \citenamefont {Wernsdorfer}, \citenamefont {Tkachuk}, \citenamefont
  {Mailly},\ and\ \citenamefont {Barbara}}]{giraudNuclearSpinDriven2001}%
  \BibitemOpen
  \bibfield  {author} {\bibinfo {author} {\bibfnamefont {R.}~\bibnamefont
  {Giraud}}, \bibinfo {author} {\bibfnamefont {W.}~\bibnamefont {Wernsdorfer}},
  \bibinfo {author} {\bibfnamefont {A.~M.}\ \bibnamefont {Tkachuk}}, \bibinfo
  {author} {\bibfnamefont {D.}~\bibnamefont {Mailly}},\ and\ \bibinfo {author}
  {\bibfnamefont {B.}~\bibnamefont {Barbara}},\ }\bibfield  {title} {\bibinfo
  {title} {Nuclear {{Spin Driven Quantum Relaxation}} in
  {{LiY}}{\textsubscript{0.998}}{{Ho}}{\textsubscript{0.002}}{{F}}{\textsubscript{4}}},\
  }\href {https://doi.org/10.1103/PhysRevLett.87.057203} {\bibfield  {journal}
  {\bibinfo  {journal} {Physical Review Letters}\ }\textbf {\bibinfo {volume}
  {87}},\ \bibinfo {pages} {057203} (\bibinfo {year} {2001})}\BibitemShut
  {NoStop}%
\bibitem [{\citenamefont {Ghosh}\ \emph {et~al.}(2002)\citenamefont {Ghosh},
  \citenamefont {Parthasarathy}, \citenamefont {Rosenbaum},\ and\ \citenamefont
  {Aeppli}}]{ghoshCoherentSpinOscillations2002}%
  \BibitemOpen
  \bibfield  {author} {\bibinfo {author} {\bibfnamefont {S.}~\bibnamefont
  {Ghosh}}, \bibinfo {author} {\bibfnamefont {R.}~\bibnamefont
  {Parthasarathy}}, \bibinfo {author} {\bibfnamefont {T.~F.}\ \bibnamefont
  {Rosenbaum}},\ and\ \bibinfo {author} {\bibfnamefont {G.}~\bibnamefont
  {Aeppli}},\ }\bibfield  {title} {\bibinfo {title} {Coherent {{Spin
  Oscillations}} in a {{Disordered Magnet}}},\ }\href
  {https://doi.org/10.1126/science.1070731} {\bibfield  {journal} {\bibinfo
  {journal} {Science}\ }\textbf {\bibinfo {volume} {296}},\ \bibinfo {pages}
  {2195} (\bibinfo {year} {2002})}\BibitemShut {NoStop}%
\bibitem [{\citenamefont {Chakraborty}\ \emph {et~al.}(2004)\citenamefont
  {Chakraborty}, \citenamefont {Henelius}, \citenamefont {Kj{\o}nsberg},
  \citenamefont {Sandvik},\ and\ \citenamefont
  {Girvin}}]{chakrabortyTheoryMagneticPhase2004}%
  \BibitemOpen
  \bibfield  {author} {\bibinfo {author} {\bibfnamefont {P.~B.}\ \bibnamefont
  {Chakraborty}}, \bibinfo {author} {\bibfnamefont {P.}~\bibnamefont
  {Henelius}}, \bibinfo {author} {\bibfnamefont {H.}~\bibnamefont
  {Kj{\o}nsberg}}, \bibinfo {author} {\bibfnamefont {A.~W.}\ \bibnamefont
  {Sandvik}},\ and\ \bibinfo {author} {\bibfnamefont {S.~M.}\ \bibnamefont
  {Girvin}},\ }\bibfield  {title} {\bibinfo {title} {Theory of the magnetic
  phase diagram of {{Li}{Ho}{F}}\textsubscript{4}},\ }\href
  {https://doi.org/10.1103/PhysRevB.70.144411} {\bibfield  {journal} {\bibinfo
  {journal} {Physical Review B}\ }\textbf {\bibinfo {volume} {70}},\ \bibinfo
  {pages} {144411} (\bibinfo {year} {2004})}\BibitemShut {NoStop}%
\bibitem [{\citenamefont {Schechter}\ and\ \citenamefont
  {Stamp}(2005)}]{schechterSignificanceHyperfineInteractions2005}%
  \BibitemOpen
  \bibfield  {author} {\bibinfo {author} {\bibfnamefont {M.}~\bibnamefont
  {Schechter}}\ and\ \bibinfo {author} {\bibfnamefont {P.~C.~E.}\ \bibnamefont
  {Stamp}},\ }\bibfield  {title} {\bibinfo {title} {Significance of the
  {{Hyperfine Interactions}} in the {{Phase Diagram}} of
  {LiHo\textsubscript{\emph{x}}Y\textsubscript{1-\emph{x}}F\textsubscript{4}}},\
  }\href {https://doi.org/10.1103/PhysRevLett.95.267208} {\bibfield  {journal}
  {\bibinfo  {journal} {Physical Review Letters}\ }\textbf {\bibinfo {volume}
  {95}},\ \bibinfo {pages} {267208} (\bibinfo {year} {2005})}\BibitemShut
  {NoStop}%
\bibitem [{\citenamefont {Tabei}\ \emph {et~al.}(2006)\citenamefont {Tabei},
  \citenamefont {Gingras}, \citenamefont {Kao}, \citenamefont {Stasiak},\ and\
  \citenamefont {Fortin}}]{tabeiInducedRandomFields2006}%
  \BibitemOpen
  \bibfield  {author} {\bibinfo {author} {\bibfnamefont {S.~M.~A.}\
  \bibnamefont {Tabei}}, \bibinfo {author} {\bibfnamefont {M.~J.~P.}\
  \bibnamefont {Gingras}}, \bibinfo {author} {\bibfnamefont {Y.-J.}\
  \bibnamefont {Kao}}, \bibinfo {author} {\bibfnamefont {P.}~\bibnamefont
  {Stasiak}},\ and\ \bibinfo {author} {\bibfnamefont {J.-Y.}\ \bibnamefont
  {Fortin}},\ }\bibfield  {title} {\bibinfo {title} {Induced {{Random Fields}}
  in the
  {LiHo\textsubscript{\emph{x}}Y\textsubscript{1-\emph{x}}F\textsubscript{4}}
  {{Quantum Ising Magnet}} in a {{Transverse Magnetic Field}}},\ }\href
  {https://doi.org/10.1103/PhysRevLett.97.237203} {\bibfield  {journal}
  {\bibinfo  {journal} {Physical Review Letters}\ }\textbf {\bibinfo {volume}
  {97}},\ \bibinfo {pages} {237203} (\bibinfo {year} {2006})}\BibitemShut
  {NoStop}%
\bibitem [{\citenamefont {Biltmo}\ and\ \citenamefont
  {Henelius}(2007)}]{biltmoPhaseDiagramDilute2007}%
  \BibitemOpen
  \bibfield  {author} {\bibinfo {author} {\bibfnamefont {A.}~\bibnamefont
  {Biltmo}}\ and\ \bibinfo {author} {\bibfnamefont {P.}~\bibnamefont
  {Henelius}},\ }\bibfield  {title} {\bibinfo {title} {Phase diagram of the
  dilute magnet
  {LiHo\textsubscript{\emph{x}}Y\textsubscript{1-\emph{x}}F\textsubscript{4}}},\
  }\href {https://doi.org/10.1103/PhysRevB.76.054423} {\bibfield  {journal}
  {\bibinfo  {journal} {Physical Review B}\ }\textbf {\bibinfo {volume} {76}},\
  \bibinfo {pages} {054423} (\bibinfo {year} {2007})}\BibitemShut {NoStop}%
\bibitem [{\citenamefont
  {Schechter}(2008)}]{schechterLiHoRandomfieldIsing2008}%
  \BibitemOpen
  \bibfield  {author} {\bibinfo {author} {\bibfnamefont {M.}~\bibnamefont
  {Schechter}},\ }\bibfield  {title} {\bibinfo {title}
  {{LiHo\textsubscript{\emph{x}}Y\textsubscript{1-\emph{x}}F\textsubscript{4}}
  4 as a random-field {{Ising}} ferromagnet},\ }\href
  {https://doi.org/10.1103/PhysRevB.77.020401} {\bibfield  {journal} {\bibinfo
  {journal} {Physical Review B}\ }\textbf {\bibinfo {volume} {77}},\ \bibinfo
  {pages} {020401} (\bibinfo {year} {2008})}\BibitemShut {NoStop}%
\bibitem [{\citenamefont {Schechter}\ and\ \citenamefont
  {Stamp}(2008)}]{schechterDerivationLowPhase2008}%
  \BibitemOpen
  \bibfield  {author} {\bibinfo {author} {\bibfnamefont {M.}~\bibnamefont
  {Schechter}}\ and\ \bibinfo {author} {\bibfnamefont {P.~C.~E.}\ \bibnamefont
  {Stamp}},\ }\bibfield  {title} {\bibinfo {title} {Derivation of the low-
  {{T}} phase diagram of
  {LiHo\textsubscript{\emph{x}}Y\textsubscript{1-\emph{x}}F\textsubscript{4}}:
  {{A}} dipolar quantum {{Ising}} magnet},\ }\href
  {https://doi.org/10.1103/PhysRevB.78.054438} {\bibfield  {journal} {\bibinfo
  {journal} {Physical Review B}\ }\textbf {\bibinfo {volume} {78}},\ \bibinfo
  {pages} {054438} (\bibinfo {year} {2008})}\BibitemShut {NoStop}%
\bibitem [{\citenamefont {Tabei}\ \emph {et~al.}(2008)\citenamefont {Tabei},
  \citenamefont {Gingras}, \citenamefont {Kao},\ and\ \citenamefont
  {Yavors'kii}}]{tabeiPerturbativeQuantumMonte2008}%
  \BibitemOpen
  \bibfield  {author} {\bibinfo {author} {\bibfnamefont {S.~M.~A.}\
  \bibnamefont {Tabei}}, \bibinfo {author} {\bibfnamefont {M.~J.~P.}\
  \bibnamefont {Gingras}}, \bibinfo {author} {\bibfnamefont {Y.-J.}\
  \bibnamefont {Kao}},\ and\ \bibinfo {author} {\bibfnamefont {T.}~\bibnamefont
  {Yavors'kii}},\ }\bibfield  {title} {\bibinfo {title} {Perturbative quantum
  {{Monte Carlo}} study of {{LiHoF}} 4 in a transverse magnetic field},\ }\href
  {https://doi.org/10.1103/PhysRevB.78.184408} {\bibfield  {journal} {\bibinfo
  {journal} {Physical Review B}\ }\textbf {\bibinfo {volume} {78}},\ \bibinfo
  {pages} {184408} (\bibinfo {year} {2008})}\BibitemShut {NoStop}%
\bibitem [{\citenamefont {Tam}\ and\ \citenamefont
  {Gingras}(2009)}]{tamSpinGlassTransitionNonzero2009}%
  \BibitemOpen
  \bibfield  {author} {\bibinfo {author} {\bibfnamefont {K.-M.}\ \bibnamefont
  {Tam}}\ and\ \bibinfo {author} {\bibfnamefont {M.~J.~P.}\ \bibnamefont
  {Gingras}},\ }\bibfield  {title} {\bibinfo {title} {Spin-{{Glass Transition}}
  at {{Nonzero Temperature}} in a {{Disordered Dipolar Ising System}}: {{The
  Case}} of
  {${\mathrm{LiHo}}_{x}{\mathrm{Y}}_{1\ensuremath{-}x}{\mathrm{F}}_{4}$}},\
  }\href {https://doi.org/10.1103/PhysRevLett.103.087202} {\bibfield  {journal}
  {\bibinfo  {journal} {Physical Review Letters}\ }\textbf {\bibinfo {volume}
  {103}},\ \bibinfo {pages} {087202} (\bibinfo {year} {2009})}\BibitemShut
  {NoStop}%
\bibitem [{\citenamefont {Gingras}\ and\ \citenamefont
  {Henelius}(2011)}]{gingrasCollectivePhenomenaLiHo2011}%
  \BibitemOpen
  \bibfield  {author} {\bibinfo {author} {\bibfnamefont {M.~J.~P.}\
  \bibnamefont {Gingras}}\ and\ \bibinfo {author} {\bibfnamefont
  {P.}~\bibnamefont {Henelius}},\ }\bibfield  {title} {\bibinfo {title}
  {Collective {{Phenomena}} in the
  {{LiHo}}{\textsubscript{{\emph{x}}}}{{Y}}{\textsubscript{1-{\emph{x}}}}{{F}}{\textsubscript{4}}
  {{Quantum Ising Magnet}}: {{Recent Progress}} and {{Open Questions}}},\
  }\href {https://doi.org/10.1088/1742-6596/320/1/012001} {\bibfield  {journal}
  {\bibinfo  {journal} {Journal of Physics: Conference Series}\ }\textbf
  {\bibinfo {volume} {320}},\ \bibinfo {pages} {012001} (\bibinfo {year}
  {2011})}\BibitemShut {NoStop}%
\bibitem [{\citenamefont {Dollberg}\ \emph {et~al.}(2022)\citenamefont
  {Dollberg}, \citenamefont {Andresen},\ and\ \citenamefont
  {Schechter}}]{dollbergEffectIntrinsicQuantum2022b}%
  \BibitemOpen
  \bibfield  {author} {\bibinfo {author} {\bibfnamefont {T.}~\bibnamefont
  {Dollberg}}, \bibinfo {author} {\bibfnamefont {J.~C.}\ \bibnamefont
  {Andresen}},\ and\ \bibinfo {author} {\bibfnamefont {M.}~\bibnamefont
  {Schechter}},\ }\bibfield  {title} {\bibinfo {title} {Effect of intrinsic
  quantum fluctuations on the phase diagram of anisotropic dipolar magnets},\
  }\href {https://doi.org/10.1103/PhysRevB.105.L180413} {\bibfield  {journal}
  {\bibinfo  {journal} {Physical Review B}\ }\textbf {\bibinfo {volume}
  {105}},\ \bibinfo {pages} {L180413} (\bibinfo {year} {2022})}\BibitemShut
  {NoStop}%
\bibitem [{\citenamefont {Dollberg}\ and\ \citenamefont
  {Schechter}(2024)}]{dollbergLiHoF4SpinhalfNonstandard2024}%
  \BibitemOpen
  \bibfield  {author} {\bibinfo {author} {\bibfnamefont {T.}~\bibnamefont
  {Dollberg}}\ and\ \bibinfo {author} {\bibfnamefont {M.}~\bibnamefont
  {Schechter}},\ }\bibfield  {title} {\bibinfo {title}
  {{{LiHoF\textsubscript{4}}} as a spin-half non-standard quantum {{Ising}}
  system},\ }\href {https://doi.org/10.21468/SciPostPhys.17.1.028} {\bibfield
  {journal} {\bibinfo  {journal} {SciPost Physics}\ }\textbf {\bibinfo {volume}
  {17}},\ \bibinfo {pages} {028} (\bibinfo {year} {2024})}\BibitemShut
  {NoStop}%
\bibitem [{\citenamefont {Beauvillain}\ \emph {et~al.}(1977)\citenamefont
  {Beauvillain}, \citenamefont {Renard},\ and\ \citenamefont
  {Hansen}}]{beauvillainLowtemperatureMagneticSusceptibility1977}%
  \BibitemOpen
  \bibfield  {author} {\bibinfo {author} {\bibfnamefont {P.}~\bibnamefont
  {Beauvillain}}, \bibinfo {author} {\bibfnamefont {J.-P.}\ \bibnamefont
  {Renard}},\ and\ \bibinfo {author} {\bibfnamefont {P.-E.}\ \bibnamefont
  {Hansen}},\ }\bibfield  {title} {\bibinfo {title} {Low-temperature magnetic
  susceptibility of {{LiErF\textsubscript{4}}}: Evidence of antiferromagnetic
  ordering at {{0.38K}}},\ }\href {https://doi.org/10.1088/0022-3719/10/24/007}
  {\bibfield  {journal} {\bibinfo  {journal} {Journal of Physics C: Solid State
  Physics}\ }\textbf {\bibinfo {volume} {10}},\ \bibinfo {pages} {L709}
  (\bibinfo {year} {1977})}\BibitemShut {NoStop}%
\bibitem [{\citenamefont {Piatek}\ \emph {et~al.}(2013)\citenamefont {Piatek},
  \citenamefont {Dalla~Piazza}, \citenamefont {Nikseresht}, \citenamefont
  {Tsyrulin}, \citenamefont {{\v Z}ivkovi{\'c}}, \citenamefont {Kr{\"a}mer},
  \citenamefont {Laver}, \citenamefont {Prokes}, \citenamefont {Mata{\v s}},
  \citenamefont {Christensen},\ and\ \citenamefont
  {R{\o}nnow}}]{piatekPhaseDiagramEnhanced2013}%
  \BibitemOpen
  \bibfield  {author} {\bibinfo {author} {\bibfnamefont {J.~O.}\ \bibnamefont
  {Piatek}}, \bibinfo {author} {\bibfnamefont {B.}~\bibnamefont
  {Dalla~Piazza}}, \bibinfo {author} {\bibfnamefont {N.}~\bibnamefont
  {Nikseresht}}, \bibinfo {author} {\bibfnamefont {N.}~\bibnamefont
  {Tsyrulin}}, \bibinfo {author} {\bibfnamefont {I.}~\bibnamefont {{\v
  Z}ivkovi{\'c}}}, \bibinfo {author} {\bibfnamefont {K.~W.}\ \bibnamefont
  {Kr{\"a}mer}}, \bibinfo {author} {\bibfnamefont {M.}~\bibnamefont {Laver}},
  \bibinfo {author} {\bibfnamefont {K.}~\bibnamefont {Prokes}}, \bibinfo
  {author} {\bibfnamefont {S.}~\bibnamefont {Mata{\v s}}}, \bibinfo {author}
  {\bibfnamefont {N.~B.}\ \bibnamefont {Christensen}},\ and\ \bibinfo {author}
  {\bibfnamefont {H.~M.}\ \bibnamefont {R{\o}nnow}},\ }\bibfield  {title}
  {\bibinfo {title} {Phase diagram with an enhanced spin-glass region of the
  mixed {{Ising--}}{XY} magnet
  {LiHo\textsubscript{\emph{x}}Er\textsubscript{1-\emph{x}}F\textsubscript{4}}},\
  }\href {https://doi.org/10.1103/PhysRevB.88.014408} {\bibfield  {journal}
  {\bibinfo  {journal} {Physical Review B}\ }\textbf {\bibinfo {volume} {88}},\
  \bibinfo {pages} {014408} (\bibinfo {year} {2013})}\BibitemShut {NoStop}%
\bibitem [{\citenamefont
  {Dalla~Piazza}(2009)}]{dallapiazzaMeanFieldCalculationsDiluted2009}%
  \BibitemOpen
  \bibfield  {author} {\bibinfo {author} {\bibfnamefont {B.}~\bibnamefont
  {Dalla~Piazza}},\ }\emph {\bibinfo {title} {Mean-{{Field}} Calculations on
  the Diluted Dipolar Magnet
  {LiHo\textsubscript{1-\emph{x}}Y\textsubscript{\emph{x}}F\textsubscript{4}}}},\
  \href@noop {} {\bibinfo {type} {M.Sc. thesis}},\ \bibinfo  {school} {EPFL}
  (\bibinfo {year} {2009})\BibitemShut {NoStop}%
\bibitem [{\citenamefont {Piatek}(2012)}]{piatekUltraLowTemperature2012}%
  \BibitemOpen
  \bibfield  {author} {\bibinfo {author} {\bibfnamefont {J.}~\bibnamefont
  {Piatek}},\ }\emph {\bibinfo {title} {Ultra {{Low Temperature Susceptometer}}
  and {{Magnetometer}}: {{Study}} of the {{Spin Glass Series
  {LiHo\textsubscript{\emph{x}}Er\textsubscript{1-\emph{x}}F\textsubscript{4}}}}}},\
  \href {http://infoscience.epfl.ch/record/181535} {Ph.D. thesis},\ \bibinfo
  {school} {EPFL} (\bibinfo {year} {2012})\BibitemShut {NoStop}%
\bibitem [{\citenamefont {Wang}\ and\ \citenamefont
  {Holm}(2001)}]{wangEstimateCutoffErrors2001}%
  \BibitemOpen
  \bibfield  {author} {\bibinfo {author} {\bibfnamefont {Z.}~\bibnamefont
  {Wang}}\ and\ \bibinfo {author} {\bibfnamefont {C.}~\bibnamefont {Holm}},\
  }\bibfield  {title} {\bibinfo {title} {Estimate of the cutoff errors in the
  {{Ewald}} summation for dipolar systems},\ }\href
  {https://doi.org/10.1063/1.1398588} {\bibfield  {journal} {\bibinfo
  {journal} {The Journal of Chemical Physics}\ }\textbf {\bibinfo {volume}
  {115}},\ \bibinfo {pages} {6351} (\bibinfo {year} {2001})}\BibitemShut
  {NoStop}%
\bibitem [{\citenamefont {Hansen}\ \emph {et~al.}(1975)\citenamefont {Hansen},
  \citenamefont {Johansson},\ and\ \citenamefont
  {Nevald}}]{hansenMagneticPropertiesLithium1975}%
  \BibitemOpen
  \bibfield  {author} {\bibinfo {author} {\bibfnamefont {P.~E.}\ \bibnamefont
  {Hansen}}, \bibinfo {author} {\bibfnamefont {T.}~\bibnamefont {Johansson}},\
  and\ \bibinfo {author} {\bibfnamefont {R.}~\bibnamefont {Nevald}},\
  }\bibfield  {title} {\bibinfo {title} {Magnetic properties of lithium
  rare-earth fluorides: {{Ferromagnetism}} in {{LiErF}}\textsubscript{4} and
  {{LiHoF}}\textsubscript{4} and crystal-field parameters at the rare-earth and
  {{Li}} sites},\ }\href {https://doi.org/10.1103/PhysRevB.12.5315} {\bibfield
  {journal} {\bibinfo  {journal} {Physical Review B}\ }\textbf {\bibinfo
  {volume} {12}},\ \bibinfo {pages} {5315} (\bibinfo {year}
  {1975})}\BibitemShut {NoStop}%
\bibitem [{\citenamefont {Magari{\~n}o}\ \emph {et~al.}(1980)\citenamefont
  {Magari{\~n}o}, \citenamefont {Tuchendler}, \citenamefont {Beauvillain},\
  and\ \citenamefont {Laursen}}]{magarinoEPRExperimentsLiTb1980}%
  \BibitemOpen
  \bibfield  {author} {\bibinfo {author} {\bibfnamefont {J.}~\bibnamefont
  {Magari{\~n}o}}, \bibinfo {author} {\bibfnamefont {J.}~\bibnamefont
  {Tuchendler}}, \bibinfo {author} {\bibfnamefont {P.}~\bibnamefont
  {Beauvillain}},\ and\ \bibinfo {author} {\bibfnamefont {I.}~\bibnamefont
  {Laursen}},\ }\bibfield  {title} {\bibinfo {title} {{{EPR}} experiments in
  {{LiTb}}{${\mathrm{F}}_{4}$}, {{LiHo}}{${\mathrm{F}}_{4}$}, and
  {{LiEr}}{${\mathrm{F}}_{4}$} at submillimeter frequencies},\ }\href
  {https://doi.org/10.1103/PhysRevB.21.18} {\bibfield  {journal} {\bibinfo
  {journal} {Physical Review B}\ }\textbf {\bibinfo {volume} {21}},\ \bibinfo
  {pages} {18} (\bibinfo {year} {1980})}\BibitemShut {NoStop}%
\bibitem [{\citenamefont {Bitko}\ \emph {et~al.}(1996)\citenamefont {Bitko},
  \citenamefont {Rosenbaum},\ and\ \citenamefont
  {Aeppli}}]{bitkoQuantumCriticalBehavior1996}%
  \BibitemOpen
  \bibfield  {author} {\bibinfo {author} {\bibfnamefont {D.}~\bibnamefont
  {Bitko}}, \bibinfo {author} {\bibfnamefont {T.~F.}\ \bibnamefont
  {Rosenbaum}},\ and\ \bibinfo {author} {\bibfnamefont {G.}~\bibnamefont
  {Aeppli}},\ }\bibfield  {title} {\bibinfo {title} {Quantum {{Critical
  Behavior}} for a {{Model Magnet}}},\ }\href
  {https://doi.org/10.1103/PhysRevLett.77.940} {\bibfield  {journal} {\bibinfo
  {journal} {Physical Review Letters}\ }\textbf {\bibinfo {volume} {77}},\
  \bibinfo {pages} {940} (\bibinfo {year} {1996})}\BibitemShut {NoStop}%
\bibitem [{\citenamefont {Hukushima}\ and\ \citenamefont
  {Nemoto}(1996)}]{hukushimaExchangeMonteCarlo1996}%
  \BibitemOpen
  \bibfield  {author} {\bibinfo {author} {\bibfnamefont {K.}~\bibnamefont
  {Hukushima}}\ and\ \bibinfo {author} {\bibfnamefont {K.}~\bibnamefont
  {Nemoto}},\ }\bibfield  {title} {\bibinfo {title} {Exchange monte carlo
  method and application to spin glass simulations},\ }\href
  {https://doi.org/10.1143/JPSJ.65.1604} {\bibfield  {journal} {\bibinfo
  {journal} {Journal of the Physical Society of Japan}\ }\textbf {\bibinfo
  {volume} {65}},\ \bibinfo {pages} {1604} (\bibinfo {year}
  {1996})}\BibitemShut {NoStop}%
\bibitem [{\citenamefont {Newman}\ and\ \citenamefont
  {Barkema}(1999)}]{newmanMonteCarloMethods1999}%
  \BibitemOpen
  \bibfield  {author} {\bibinfo {author} {\bibfnamefont {M.}~\bibnamefont
  {Newman}}\ and\ \bibinfo {author} {\bibfnamefont {G.}~\bibnamefont
  {Barkema}},\ }\href {https://books.google.co.il/books?id=KKL2nQEACAAJ} {\emph
  {\bibinfo {title} {Monte Carlo Methods in Statistical Physics}}}\ (\bibinfo
  {publisher} {{Clarendon Press}},\ \bibinfo {year} {1999})\BibitemShut
  {NoStop}%
\bibitem [{\citenamefont {Berg}(2004)}]{bergMarkovChainMonte2004}%
  \BibitemOpen
  \bibfield  {author} {\bibinfo {author} {\bibfnamefont {B.~A.}\ \bibnamefont
  {Berg}},\ }\href@noop {} {\emph {\bibinfo {title} {Markov Chain {{Monte
  Carlo}} Simulations and Their Statistical Analysis: With Web-Based
  {{Fortran}} Code}}}\ (\bibinfo  {publisher} {{World Scientific}},\ \bibinfo
  {address} {{Hackensack, NJ}},\ \bibinfo {year} {2004})\BibitemShut {NoStop}%
\bibitem [{\citenamefont
  {McKenzie}(2016)}]{mckenzieFluctuationsPhaseTransitions2016}%
  \BibitemOpen
  \bibfield  {author} {\bibinfo {author} {\bibfnamefont {R.}~\bibnamefont
  {McKenzie}},\ }\emph {\bibinfo {title} {Fluctuations and phase transitions in
  quantum Ising systems}},\ \href
  {https://doi.org/http://dx.doi.org/10.14288/1.0314166} {Ph.D. thesis},\
  \bibinfo  {school} {University of British Columbia} (\bibinfo {year}
  {2016})\BibitemShut {NoStop}%
\bibitem [{\citenamefont {Katzgraber}\ \emph {et~al.}(2006)\citenamefont
  {Katzgraber}, \citenamefont {K{\"o}rner},\ and\ \citenamefont
  {Young}}]{katzgraberUniversalityThreedimensionalIsing2006}%
  \BibitemOpen
  \bibfield  {author} {\bibinfo {author} {\bibfnamefont {H.~G.}\ \bibnamefont
  {Katzgraber}}, \bibinfo {author} {\bibfnamefont {M.}~\bibnamefont
  {K{\"o}rner}},\ and\ \bibinfo {author} {\bibfnamefont {A.~P.}\ \bibnamefont
  {Young}},\ }\bibfield  {title} {\bibinfo {title} {Universality in
  three-dimensional {{Ising}} spin glasses: {{A Monte Carlo}} study},\ }\href
  {https://doi.org/10.1103/PhysRevB.73.224432} {\bibfield  {journal} {\bibinfo
  {journal} {Physical Review B}\ }\textbf {\bibinfo {volume} {73}},\ \bibinfo
  {pages} {224432} (\bibinfo {year} {2006})}\BibitemShut {NoStop}%
\bibitem [{\citenamefont {Ballesteros}\ \emph {et~al.}(2000)\citenamefont
  {Ballesteros}, \citenamefont {Cruz}, \citenamefont {Fern{\'a}ndez},
  \citenamefont {{Mart{\'i}n-Mayor}}, \citenamefont {Pech}, \citenamefont
  {{Ruiz-Lorenzo}}, \citenamefont {Taranc{\'o}n}, \citenamefont {T{\'e}llez},
  \citenamefont {Ullod},\ and\ \citenamefont
  {Ungil}}]{ballesterosCriticalBehaviorThreedimensional2000a}%
  \BibitemOpen
  \bibfield  {author} {\bibinfo {author} {\bibfnamefont {H.~G.}\ \bibnamefont
  {Ballesteros}}, \bibinfo {author} {\bibfnamefont {A.}~\bibnamefont {Cruz}},
  \bibinfo {author} {\bibfnamefont {L.~A.}\ \bibnamefont {Fern{\'a}ndez}},
  \bibinfo {author} {\bibfnamefont {V.}~\bibnamefont {{Mart{\'i}n-Mayor}}},
  \bibinfo {author} {\bibfnamefont {J.}~\bibnamefont {Pech}}, \bibinfo {author}
  {\bibfnamefont {J.~J.}\ \bibnamefont {{Ruiz-Lorenzo}}}, \bibinfo {author}
  {\bibfnamefont {A.}~\bibnamefont {Taranc{\'o}n}}, \bibinfo {author}
  {\bibfnamefont {P.}~\bibnamefont {T{\'e}llez}}, \bibinfo {author}
  {\bibfnamefont {C.~L.}\ \bibnamefont {Ullod}},\ and\ \bibinfo {author}
  {\bibfnamefont {C.}~\bibnamefont {Ungil}},\ }\bibfield  {title} {\bibinfo
  {title} {Critical behavior of the three-dimensional {{Ising}} spin glass},\
  }\href {https://doi.org/10.1103/PhysRevB.62.14237} {\bibfield  {journal}
  {\bibinfo  {journal} {Physical Review B}\ }\textbf {\bibinfo {volume} {62}},\
  \bibinfo {pages} {14237} (\bibinfo {year} {2000})}\BibitemShut {NoStop}%
\bibitem [{\citenamefont {Harada}(2011)}]{haradaBayesianInferenceScaling2011a}%
  \BibitemOpen
  \bibfield  {author} {\bibinfo {author} {\bibfnamefont {K.}~\bibnamefont
  {Harada}},\ }\bibfield  {title} {\bibinfo {title} {Bayesian inference in the
  scaling analysis of critical phenomena},\ }\href
  {https://doi.org/10.1103/PhysRevE.84.056704} {\bibfield  {journal} {\bibinfo
  {journal} {Physical Review E}\ }\textbf {\bibinfo {volume} {84}},\ \bibinfo
  {pages} {056704} (\bibinfo {year} {2011})}\BibitemShut {NoStop}%
\bibitem [{\citenamefont {Harada}(2015)}]{haradaKernelMethodCorrections2015}%
  \BibitemOpen
  \bibfield  {author} {\bibinfo {author} {\bibfnamefont {K.}~\bibnamefont
  {Harada}},\ }\bibfield  {title} {\bibinfo {title} {Kernel method for
  corrections to scaling},\ }\href {https://doi.org/10.1103/PhysRevE.92.012106}
  {\bibfield  {journal} {\bibinfo  {journal} {Physical Review E}\ }\textbf
  {\bibinfo {volume} {92}},\ \bibinfo {pages} {012106} (\bibinfo {year}
  {2015})}\BibitemShut {NoStop}%
\bibitem [{\citenamefont {Babkevich}\ \emph {et~al.}(2016)\citenamefont
  {Babkevich}, \citenamefont {Nikseresht}, \citenamefont {Kovacevic},
  \citenamefont {Piatek}, \citenamefont {Dalla~Piazza}, \citenamefont
  {Kraemer}, \citenamefont {Kr{\"a}mer}, \citenamefont {Proke{\v s}},
  \citenamefont {Mat'a{\v s}}, \citenamefont {Jensen},\ and\ \citenamefont
  {R{\o}nnow}}]{babkevichPhaseDiagramDiluted2016}%
  \BibitemOpen
  \bibfield  {author} {\bibinfo {author} {\bibfnamefont {P.}~\bibnamefont
  {Babkevich}}, \bibinfo {author} {\bibfnamefont {N.}~\bibnamefont
  {Nikseresht}}, \bibinfo {author} {\bibfnamefont {I.}~\bibnamefont
  {Kovacevic}}, \bibinfo {author} {\bibfnamefont {J.~O.}\ \bibnamefont
  {Piatek}}, \bibinfo {author} {\bibfnamefont {B.}~\bibnamefont
  {Dalla~Piazza}}, \bibinfo {author} {\bibfnamefont {C.}~\bibnamefont
  {Kraemer}}, \bibinfo {author} {\bibfnamefont {K.~W.}\ \bibnamefont
  {Kr{\"a}mer}}, \bibinfo {author} {\bibfnamefont {K.}~\bibnamefont {Proke{\v
  s}}}, \bibinfo {author} {\bibfnamefont {S.}~\bibnamefont {Mat'a{\v s}}},
  \bibinfo {author} {\bibfnamefont {J.}~\bibnamefont {Jensen}},\ and\ \bibinfo
  {author} {\bibfnamefont {H.~M.}\ \bibnamefont {R{\o}nnow}},\ }\bibfield
  {title} {\bibinfo {title} {Phase diagram of diluted {{Ising}} ferromagnet
  {{LiHo}}\textsubscript{x}{{Y}}\textsubscript{1-x}{{F}}\textsubscript{4}},\
  }\href {https://doi.org/10.1103/PhysRevB.94.174443} {\bibfield  {journal}
  {\bibinfo  {journal} {Physical Review B}\ }\textbf {\bibinfo {volume} {94}},\
  \bibinfo {pages} {174443} (\bibinfo {year} {2016})}\BibitemShut {NoStop}%
\bibitem [{\citenamefont
  {Piatek}(2009)}]{piatekCharacterisationLiHoxEr1XF42009}%
  \BibitemOpen
  \bibfield  {author} {\bibinfo {author} {\bibfnamefont {J.}~\bibnamefont
  {Piatek}},\ }\emph {\bibinfo {title} {Characterisation of
  {LiHo\textsubscript{\emph{x}}Er\textsubscript{1-\emph{x}}F\textsubscript{4}}
  Alloyed {{Model Magnets}}}},\ \href@noop {} {\bibinfo {type} {M.Sc. thesis}},\
  \bibinfo  {school} {EPFL} (\bibinfo {year} {2009})\BibitemShut {NoStop}%
\bibitem [{\citenamefont {Harris}(1974)}]{harrisEffectRandomDefects1974}%
  \BibitemOpen
  \bibfield  {author} {\bibinfo {author} {\bibfnamefont {A.~B.}\ \bibnamefont
  {Harris}},\ }\bibfield  {title} {\bibinfo {title} {Effect of random defects
  on the critical behaviour of {{Ising}} models},\ }\href
  {https://doi.org/10.1088/0022-3719/7/9/009} {\bibfield  {journal} {\bibinfo
  {journal} {Journal of Physics C: Solid State Physics}\ }\textbf {\bibinfo
  {volume} {7}},\ \bibinfo {pages} {1671} (\bibinfo {year} {1974})}\BibitemShut
  {NoStop}%
\bibitem [{\citenamefont {Ballesteros}\ \emph {et~al.}(1998)\citenamefont
  {Ballesteros}, \citenamefont {Fern{\'a}ndez}, \citenamefont
  {{Mart{\'i}n-Mayor}}, \citenamefont {Mu{\~n}oz~Sudupe}, \citenamefont
  {Parisi},\ and\ \citenamefont
  {{Ruiz-Lorenzo}}}]{ballesterosCriticalExponentsThreedimensional1998}%
  \BibitemOpen
  \bibfield  {author} {\bibinfo {author} {\bibfnamefont {H.~G.}\ \bibnamefont
  {Ballesteros}}, \bibinfo {author} {\bibfnamefont {L.~A.}\ \bibnamefont
  {Fern{\'a}ndez}}, \bibinfo {author} {\bibfnamefont {V.}~\bibnamefont
  {{Mart{\'i}n-Mayor}}}, \bibinfo {author} {\bibfnamefont {A.}~\bibnamefont
  {Mu{\~n}oz~Sudupe}}, \bibinfo {author} {\bibfnamefont {G.}~\bibnamefont
  {Parisi}},\ and\ \bibinfo {author} {\bibfnamefont {J.~J.}\ \bibnamefont
  {{Ruiz-Lorenzo}}},\ }\bibfield  {title} {\bibinfo {title} {Critical exponents
  of the three-dimensional diluted {{Ising}} model},\ }\href
  {https://doi.org/10.1103/PhysRevB.58.2740} {\bibfield  {journal} {\bibinfo
  {journal} {Physical Review B}\ }\textbf {\bibinfo {volume} {58}},\ \bibinfo
  {pages} {2740} (\bibinfo {year} {1998})}\BibitemShut {NoStop}%
\bibitem [{\citenamefont {Calabrese}\ \emph {et~al.}(2003)\citenamefont
  {Calabrese}, \citenamefont {{Mart{\'i}n-Mayor}}, \citenamefont {Pelissetto},\
  and\ \citenamefont {Vicari}}]{calabreseThreedimensionalRandomlyDilute2003}%
  \BibitemOpen
  \bibfield  {author} {\bibinfo {author} {\bibfnamefont {P.}~\bibnamefont
  {Calabrese}}, \bibinfo {author} {\bibfnamefont {V.}~\bibnamefont
  {{Mart{\'i}n-Mayor}}}, \bibinfo {author} {\bibfnamefont {A.}~\bibnamefont
  {Pelissetto}},\ and\ \bibinfo {author} {\bibfnamefont {E.}~\bibnamefont
  {Vicari}},\ }\bibfield  {title} {\bibinfo {title} {Three-dimensional randomly
  dilute {{Ising}} model: {{Monte Carlo}} results},\ }\href
  {https://doi.org/10.1103/PhysRevE.68.036136} {\bibfield  {journal} {\bibinfo
  {journal} {Physical Review E}\ }\textbf {\bibinfo {volume} {68}},\ \bibinfo
  {pages} {036136} (\bibinfo {year} {2003})}\BibitemShut {NoStop}%
\bibitem [{\citenamefont {Berche}\ \emph {et~al.}(2004)\citenamefont {Berche},
  \citenamefont {Chatelain}, \citenamefont {Berche},\ and\ \citenamefont
  {Janke}}]{bercheBondDilution3D2004}%
  \BibitemOpen
  \bibfield  {author} {\bibinfo {author} {\bibfnamefont {P.~E.}\ \bibnamefont
  {Berche}}, \bibinfo {author} {\bibfnamefont {C.}~\bibnamefont {Chatelain}},
  \bibinfo {author} {\bibfnamefont {B.}~\bibnamefont {Berche}},\ and\ \bibinfo
  {author} {\bibfnamefont {W.}~\bibnamefont {Janke}},\ }\bibfield  {title}
  {\bibinfo {title} {Bond dilution in the {{3D Ising}} model: A {{Monte Carlo}}
  study},\ }\href {https://doi.org/10.1140/epjb/e2004-00141-x} {\bibfield
  {journal} {\bibinfo  {journal} {The European Physical Journal B}\ }\textbf
  {\bibinfo {volume} {38}},\ \bibinfo {pages} {463} (\bibinfo {year}
  {2004})}\BibitemShut {NoStop}%
\bibitem [{\citenamefont {Hasenbusch}\ \emph
  {et~al.}(2007{\natexlab{a}})\citenamefont {Hasenbusch}, \citenamefont
  {Toldin}, \citenamefont {Pelissetto},\ and\ \citenamefont
  {Vicari}}]{hasenbuschUniversalityClass3D2007}%
  \BibitemOpen
  \bibfield  {author} {\bibinfo {author} {\bibfnamefont {M.}~\bibnamefont
  {Hasenbusch}}, \bibinfo {author} {\bibfnamefont {F.~P.}\ \bibnamefont
  {Toldin}}, \bibinfo {author} {\bibfnamefont {A.}~\bibnamefont {Pelissetto}},\
  and\ \bibinfo {author} {\bibfnamefont {E.}~\bibnamefont {Vicari}},\
  }\bibfield  {title} {\bibinfo {title} {The universality class of {{3D}}
  site-diluted and bond-diluted {{Ising}} systems},\ }\href
  {https://doi.org/10.1088/1742-5468/2007/02/P02016} {\bibfield  {journal}
  {\bibinfo  {journal} {Journal of Statistical Mechanics: Theory and
  Experiment}\ }\textbf {\bibinfo {volume} {2007}},\ \bibinfo {pages} {P02016}
  (\bibinfo {year} {2007}{\natexlab{a}})}\BibitemShut {NoStop}%
\bibitem [{\citenamefont {Hasenbusch}\ \emph
  {et~al.}(2007{\natexlab{b}})\citenamefont {Hasenbusch}, \citenamefont
  {Toldin}, \citenamefont {Pelissetto},\ and\ \citenamefont
  {Vicari}}]{hasenbuschCriticalBehaviorThreedimensional2007}%
  \BibitemOpen
  \bibfield  {author} {\bibinfo {author} {\bibfnamefont {M.}~\bibnamefont
  {Hasenbusch}}, \bibinfo {author} {\bibfnamefont {F.~P.}\ \bibnamefont
  {Toldin}}, \bibinfo {author} {\bibfnamefont {A.}~\bibnamefont {Pelissetto}},\
  and\ \bibinfo {author} {\bibfnamefont {E.}~\bibnamefont {Vicari}},\
  }\bibfield  {title} {\bibinfo {title} {Critical behavior of the
  three-dimensional \textpm{} {{J Ising}} model at the
  paramagnetic-ferromagnetic transition line},\ }\href
  {https://doi.org/10.1103/PhysRevB.76.094402} {\bibfield  {journal} {\bibinfo
  {journal} {Physical Review B}\ }\textbf {\bibinfo {volume} {76}},\ \bibinfo
  {pages} {094402} (\bibinfo {year} {2007}{\natexlab{b}})}\BibitemShut
  {NoStop}%
\bibitem [{\citenamefont {Ronnow}\ \emph {et~al.}(2007)\citenamefont {Ronnow},
  \citenamefont {Jensen}, \citenamefont {Parthasarathy}, \citenamefont
  {Aeppli}, \citenamefont {Rosenbaum}, \citenamefont {McMorrow},\ and\
  \citenamefont {Kraemer}}]{ronnowMagneticExcitationsQuantum2007}%
  \BibitemOpen
  \bibfield  {author} {\bibinfo {author} {\bibfnamefont {H.~M.}\ \bibnamefont
  {Ronnow}}, \bibinfo {author} {\bibfnamefont {J.}~\bibnamefont {Jensen}},
  \bibinfo {author} {\bibfnamefont {R.}~\bibnamefont {Parthasarathy}}, \bibinfo
  {author} {\bibfnamefont {G.}~\bibnamefont {Aeppli}}, \bibinfo {author}
  {\bibfnamefont {T.~F.}\ \bibnamefont {Rosenbaum}}, \bibinfo {author}
  {\bibfnamefont {D.~F.}\ \bibnamefont {McMorrow}},\ and\ \bibinfo {author}
  {\bibfnamefont {C.}~\bibnamefont {Kraemer}},\ }\bibfield  {title} {\bibinfo
  {title} {Magnetic excitations near the quantum phase transition in the
  {{Ising}} ferromagnet {{LiHoF}}\textsubscript{4}},\ }\href
  {https://doi.org/10.1103/PhysRevB.75.054426} {\bibfield  {journal} {\bibinfo
  {journal} {Physical Review B}\ }\textbf {\bibinfo {volume} {75}},\ \bibinfo
  {pages} {054426} (\bibinfo {year} {2007})}\BibitemShut {NoStop}%
\end{thebibliography}
\end{document}